\documentclass[prb,aps,superscriptaddress,amsmath,amssymb,amsfonts,twocolumn,floatfix
]{revtex4-2}

\usepackage[T1]{fontenc}
\usepackage{amssymb}
\usepackage{graphicx}
\usepackage{color}
\usepackage{caption}
\usepackage{hyperref}
\usepackage{amsthm}
\usepackage{amsmath}
\usepackage{braket}
\usepackage{appendix}
\usepackage{subcaption}
\usepackage{enumitem}
\usepackage{comment}

\usepackage{tikz}
\usetikzlibrary{patterns}

\newcommand{\Mod}[1]{\ (\mathrm{mod}\ #1)}

\newcommand{\gcdM}{g }
\newtheorem{theorem}{Theorem}
\newtheorem*{theorem*}{Theorem}

\begin{document}

\title{The fate of local order in topologically frustrated spin chains}

\author{V. Mari\'c}
\email{vmaric@sissa.it}
\affiliation{Institut Ru\dj er Bo\v{s}kovi\'c, Bijeni\v{c}ka cesta 54, 10000 Zagreb, Croatia}
\affiliation{SISSA and INFN, via Bonomea 265, 34136 Trieste, Italy}

\author{S. M. Giampaolo}
\email{sgiampa@irb.hr}
\affiliation{Institut Ru\dj er Bo\v{s}kovi\'c, Bijeni\v{c}ka cesta 54, 10000 Zagreb, Croatia}

\author{Fabio Franchini}
\email{fabio.franchini@irb.hr}
\affiliation{Institut Ru\dj er Bo\v{s}kovi\'c, Bijeni\v{c}ka cesta 54, 10000 Zagreb, Croatia}

\begin{abstract}
	
It has been recently shown that the presence of topological frustration, induced by periodic boundary conditions in an antiferromagnetic $XY$ chain made of an odd number of spins, prevents the realization of a perfectly staggered local order.  
Starting from this result and exploiting a recently introduced approach which enables the direct calculation of the expectation value of any operator with support over a finite range of lattice sites, in this work we investigate the possible fates of local orders. 
We show that, regardless of the variety of possible situations, they can be all arranged in two different cases.
A system admits a finite local order only if the ground state is degenerate, with at least two elements whose momenta differ, in the thermodynamic limit, by $\pi$, and this order breaks translational symmetry.
In all other cases, any local order decays to zero, algebraically (or faster) in the chain length.
Moreover, we show that, in some cases, which of the two possibilities is realized, may depend on the sequence of chain lengths with which the thermodynamic limit is reached.
These results are established both analytically and by exact diagonalization and illustrated through examples.
\end{abstract}

\maketitle

\section{Introduction}

Frustration arises as a competition between terms promoting incompatible arrangements. 
Although this definition applies also to quantum Hamiltonians with unfrustrated
counterpart~\cite{Wolf03,Giampaolo2011,Marzolino13}, it is usually meant in its classical origin, known as {\it geometrical frustration}~\cite{Toulouse77,Vannimenus77}. 
Usually, in frustrated systems, one can identify several frustrated loops, either induced by competing long-range terms or just because of the lattice geometry. 
In such cases, the amount of frustration scales with the system's size and the interplay between local interactions, quantum effects, and the non-local nature of geometrical frustration renders the study of these systems very challenging.
On the other side, their phenomenology is very rich, displaying algebraic decays of correlation functions not associated to criticality~\cite{Huse03, Henley05}, localized zero energy modes~\cite{Villain79, Ritchey93, Oleg2002, Lee02}, non-zero entropy at near-zero temperature~\cite{Harris97, Ramirez99}, etc.
Due to this fact, they are also platforms to realize interesting emergent properties, such as artificial electromagnetism~\cite{Huse03, Henley05} monopoles and Dirac strings~\cite{Morris09}. 
Moreover, magnetic frustrated systems are among the best candidates to host the elusive spin liquid phase~\cite{Balents10}. 

However, differently from what was expected, it has been recently realized that even simple systems, with a much weaker degree of frustration, namely with a number of frustrated loops that does not scale with the system size, can host surprises.
This is the case of systems with a short-range antiferromagnetic interaction in which a staggered arrangement is made impossible by the assumption of Frustrated Boundary Conditions (FBC), i.e. periodic boundary conditions applied on a chain made of an odd number of sites.
Classically, such frustration produces a massive degeneracy in the lowest energy state, because each such state develops a domain wall defect, which can be located at any site of the chain. 
Quantum interactions lift this degeneracy to a band of states, which can largely be characterized as states with a single traveling excitation. 
While these aspects have been understood qualitatively a long time ago, only lately their quantitative appraisal revealed their deep consequences.

First, it has been found that the antiferromagnetic systems with FBC are gapless, with non-relativistic gapless excitations~\cite{Laumann2012, Cabrera1986, Cabrera1987, Barber1987}. 
Then, it has been established that perfect FBC constitute a quantum phase transition point with respect to different boundary conditions~\cite{CampostriniRings}, that the spin-correlation functions at large distances develop unusual algebraic corrections~\cite{Dong2016}, and that the entanglement entropy in the ground state indeed carries the signature of a single excitation over the ground state~\cite{Giampaolo2018}. 
More importantly, it has been shown in~\cite{Maric2019, MaricToeplitz} that the topological frustration that characterizes such systems can destroy the order parameter. 
The traditional order is staggered and quantum interactions resolve the conflict between it and the FBC with an interference pattern that effectively cancels the magnetization, leaving only a mesoscopic ferromagnetic order at finite sizes, that vanishes algebraically with the chain length. 
This phenomenology has later been enriched.
Indeed, it was found~\cite{Maric2020CP} that a different interference pattern, allowed by an enlarged ground state manifold degeneracy, can also admit an incommensurate antiferromagnet, characterized by a magnetization profile that varies in space with an incommensurate pattern. 
This type of order has later been shown to be stable against antiferromagnetic (AFM) defects~\cite{Torre2021}. 
Moreover, the boundary between the mesoscopic ferromagnetic order and the incommensurate AFM one is a first-order quantum phase transition, which exists only in presence of FBC~\cite{Maric2020CP}.
All these results have established that, contrary to standard expectations, the boundary conditions can indeed affect the local, bulk behavior of a system, or, at least, that this is the case in presence of frustration, opening a gateway to connect the physics of simply frustrated chains to that of generic frustrated systems. 

It should, however, be remarked that the results discussed above have been found in specific (integrable) models and one should wonder about their general relevance. 
In this work, we address the question of whether topological frustration generically destroys local order or creates a modulated AFM order with a site-dependent magnetization.
As it is well known, local order parameters are central elements in Ginzburg-Landau theory. 
They are expectation values of local operators, with support over a finite range of lattice sites, which, given the symmetries of the system, should vanish and which, when they assume a value other than zero, signal the spontaneous breaking of the symmetry and the establishment of a macroscopic order.
We consider general spin-1/2 models with a dominant short-range antiferromagnetic interaction, subject only to the symmetry constraint that their Hamiltonians do not change under spatial translation and commute with the parity operators in all three spin directions.
As a matter of fact, these assumptions apply to a wide class of systems without external fields and defects, including ones with short- and long-range two-body Ising-like interactions, cluster terms etc.
When the lattice has an odd number of sites, the property to commutativity with the parity operators ensures the presence of an exact (Kramers) degeneracy in the ground state manifold, which is always spanned by an even number of states. This allows for the direct evaluation, even in a finite-size system, of the expectation values for all local operators, i.e. operators with support over a finite range of lattice sites~\cite{Maric2019,Maric2020CP}. 
One can then follow the behavior of these observables toward the thermodynamic limit.

In this way, we show that two main pictures can be realized:
\begin{itemize}
\item If the model has at least a four fold degenerate ground state manifold with two ground-states whose momenta differ by $\pi$ in the thermodynamic limit, an incommensurate AFM order like that found in~\cite{Maric2020CP} can emerge. This solution can be interpreted as a distortion of the normal antiferromagnetic order created by the system in order to adapt to the FBC.
Indeed, in this way the system preserves a semblance of the usual order but with a modulation over the whole chain, which spontaneously breaks translational invariance. 
; 
\item On the contrary, if there are not two ground states whose momenta differ by $\pi$ in the thermodynamic limit, then any expectation value that can play the role of local order parameter decays algebraically (or faster) to zero with the system size.
This case can be separated in two sub-cases. 
If the system admits a two-fold degenerate ground-state, each one of them has a zero momentum and the only possible local order is a mesoscopic non-staggered ferromagnetic order~\cite{Maric2019}.
On the contrary, if the ground-state manifold has a dimension greater than two then the system can show both ferromagnetic and incommensurate-staggered mesoscopic magnetization patterns~\cite{Maric2021modify}.
\end{itemize}

In particular, these results imply that, when the boundary conditions kill the order parameter connected to the dominant interaction (namely, the magnetization), these systems are unable to develop any other type of order with support over a finite range of lattice sites, regardless of the type and nature of the other interactions in the Hamiltonian.

To determine the ground state properties and analyze the local order in generic systems, we will take advantage of the Hilbert space structure at a classical point (with simple domain wall as lowest energy states) and use a highly degenerate perturbation theory. 
This will allow us to classify whether in a finite neighborhood of the classical point any order vanishes in the thermodynamic limit or if a finite incommensurate order can emerge.
While the amplitude of any order generally depends on the microscopic details of the model, its finiteness is a property of the given phase and thus to establish its existence (or lack thereof) it is sufficient just to consider a small finite parameter region.
We will also corroborate these findings through the exact numerical diagonalization of a few examples, as well as the analytical solution of a series of Cluster-Ising models that showcase various phenomenologies.

The paper is organized as following: we start by lying the foundations and notations for our analysis in Sec.~\ref{sec:AntiCommSym} and \ref{sec:TranslSymm}, by discussing the importance and implications of the symmetries that lead to the Kramers degeneracy and the general structure of the ground states for the models we consider. Then, in Sec.~\ref{sec:MatrixElements} we present our main results in the form of two theorems that provide bounds on the matrix elements of local operators. These results are used in Sec.~\ref{sec:LocalOrder} to explain what types of order are possible in chains with FBC, while Sec.~\ref{sec:Examples} contains a few relevant examples to clarify our analysis: in Sec.~\ref{sec:2body} chains with pure 2-body interactions (even beyond nearest neighbor) are considered, while in Sec.~\ref{sec:ClusterIsing} we also allow for cluster interactions. All examples are corroborated by numerical results based on exact diagonalization. Finally, Sec.~\ref{sec:Conclusions} collects our concluding remarks. We moved the technical aspects of the proofs of the two theorems in App.~\ref{Appendix Proof} and \ref{Appendix Proof2}, while App.~\ref{ClusterIsingApp} contains a details analysis of generic Cluster-Ising chains with FBC and proves their peculiar ground state degeneracy structure.

\section{Anticommuting Parity Symmetries}
\label{sec:AntiCommSym}

All along our work we focus on one-dimensional translational invariant spin-$1/2$ systems which Hamiltonians show a dominant antiferromagnetic Ising interaction in one direction, which, without loos of generality, we set to be $x$.
Together with such dominant term, the Hamiltonians are also characterized by one or more sub-dominant spatially-invariant terms so that all Hamiltonians can be written as
\begin{equation}\label{Hamiltonian AFM}
H=\sum_{j=1}^{N}\sigma_{j}^x\sigma_{j+1}^x + \lambda \sum_{j=1}^{N}H_{j} \; .
\end{equation}
Here $\sigma_j^\alpha$, for $\alpha=x,y,z$, are Pauli spin operators, the terms $H_j$ describe the sub-dominant interactions (the index $j$ is shifted to ensure translational invariance), $\lambda$ is the  relative weight of the sub-dominant term and, since the Ising term is the dominant one, we assume that $|\lambda|<1$. 
We assume that $H$ commutes with all three parity operators $\Pi^\alpha \equiv \bigotimes_{j=1}^N \sigma_j^\alpha$ ($[H,\Pi^\alpha]=0$ for $\alpha=x,y,z$), so that the whole Hamiltonian becomes invariant under transformations $\sigma_j^\alpha\to -\sigma_j^\alpha$ $\forall j$. 
Let us now consider that our system holds FBC, i.e. it has periodic boundary conditions ($\sigma_{j}^\alpha=\sigma_{j+N}^{\alpha}$) and it is made by an odd number $N$ of spins.
On a system made of an odd number of sites $N$, the three parity operators $\Pi^\alpha$ do not commute. 
Instead, they anticommute ($\left\{ \Pi^\alpha, \Pi^\beta \right\} = 2 \delta_{\alpha,\beta}$) and realize a non-local $SU(2)$ algebra ($\left[ \Pi^\alpha, \Pi^\beta \right]=\imath\,\varepsilon_{\alpha,\beta,\gamma} 2 (-1)^\frac{N-1}{2} \Pi^\gamma$).
Since the Hamiltonian \eqref{Hamiltonian AFM} commutes with all $\Pi^\alpha$, its ground state manifold is at-least two-fold degenerate~\cite{Maric2019,Maric2020CP} (which is an instance of Kramers degeneracy), and any ground state breaks at least one of the parity symmetries. 
Thus, in such a setting, to study the behavior of the order parameters in the thermodynamic limits we can explicitly evaluate it at fixed $N$ and then let $N$  diverge, avoiding the complications of the usual procedure of applying a symmetry-breaking field and removing it only after the thermodynamic limit. 

Moreover, the same structure also allows for the direct computation of matrix elements between states with different parities, whose calculation usually either requires extremely cumbersome expressions of limited practical use or is achieved indirectly from certain expectation values by invoking the cluster decomposition property.
In particular, let $\ket{g}$ be an eigenstate of $H$ and, simultaneously, an eigenstate of $\Pi^x$ with eigenvalue equal to one, i.e. $\Pi^x\ket{g}=\ket{g}$. 
Since the parity operators mutually anticommute ($\{\Pi^\alpha, \Pi^\beta\}=2\delta_{\alpha,\beta} $), it follows that the state $\Pi^z\ket{g}$ has the same energy but opposite parity with respect to $\Pi^x$, i.e. $\Pi^x\Pi^z\ket{g}=-\Pi^z\ket{g}$. 
States with different parities can be constructed through superpositions of states above and thus it is possible to calculate the ground state expectation value of operators ${\cal O}$ breaking one symmetry of the Hamiltonian by choosing a suitable ground state. 
For instance, for an eigenstate $\ket{g}$ of $\Pi^x$, the magnetization in the $x$ direction can be calculated as $\bra{g} \sigma_j^x \ket{g}$.
On the other hand, the magnetization in the $z$ direction can be evaluated on the state $\ket{\tilde g}=\frac{1}{\sqrt{2}}(\mathbf{1}+\Pi^z)\ket{g}$ and is equal to $\bra{\tilde g}\sigma^z_j\ket{\tilde g} = \bra{g} \sigma_j^z \Pi^z \ket{g}$.



\section{Translational Symmetry and the ground-states structure}
\label{sec:TranslSymm}

Let us examine the structure of the ground-states of the studied systems on the basis of general arguments. 
At the classical point $\lambda=0$ the topological frustration does not allow for every spin to point oppositely to its nearest neighbors. 
Instead, the ground space is $2N$-fold degenerate, spanned by the "kink states",
which have a single ferromagnetic bond (two spins aligned in the same direction), i.e. the "kink", and $N-1$ antiferromagnetic bonds (spins aligned in opposite directions). 
We denote by $\ket{j}$ the kink state in which the ferromagnetic bond is between sites $j$ and $j+1$, with $\bra{j}\sigma_j^x\ket{j}=1$, while the kink state that we obtain flipping all the spins, i.e. $\Pi^z\ket{j}$, has $\bra{j}\Pi^z\sigma_j^x\Pi^z\ket{j}=-1$. 
Above the states with a single kink there is an energy gap of order one separating them from the states with three kinks (due to odd $N$ an even number of kinks is not allowed). 
At higher energies, one finds bands with a progressively growing number of kinks separated by gaps of the same order as the first.

By turning on a small coupling $\lambda$ in eq.~\eqref{Hamiltonian AFM}, the degenerate states typically split in energy. 
For small $\lambda$ (much smaller than the gap between the two lowest energy bands), the ground state will be described accurately within the single kink subspace. 
Assuming, thus, $|\lambda|\ll 1$ and neglecting, for the moment, the states with more kinks, because of translational invariance we write the ground states as
\begin{equation}\label{states s p}
\!\!\!\ket{s_p} \!\equiv\! \frac{1}{\sqrt{N}}\sum\limits_{j=1}^N \! e^{\imath p j}\ket{j}, \quad \!\Pi^z\ket{s_p} \!=\! \frac{1}{\sqrt{N}}\sum\limits_{j=1}^N \! e^{\imath p j}\Pi^z\ket{j}\!.
\end{equation}
Here $p= 2\pi k /N$, with $k$ running over integers from $0$ to $N-1$, is the lattice momentum, whose quantization is a result of periodic boundary conditions. 

Increasing $\lambda$, the ground state will acquire contributions from states with more kinks but, because of translational invariance, the states can still be labeled by their momentum $p$. 
To describe the structure of such states let us introduce the translation operator $T$, a unitary operator that shifts cyclically the spins by one lattice site, i.e. $T^\dagger \sigma_j^\alpha T=\sigma_{j+1}^\alpha$, for $\alpha=x,y,z$. 
The eigenvalues $e^{\imath p}$ of the translation operator fall on the unit circle, where the angle $p$ defines the momentum of the state.
Now, for any eigenstate of the model with momentum $p$, ground state in particular, the contributions coming from the subspaces with different number of kinks can be separated. 
To show this fact let us define the state $\ket{\boldsymbol{\beta}}$ as the tensor product, on all the spins of the chain, of one of the two eigenstates of $\sigma^x_j$, i.e. 
$\ket{\boldsymbol{\beta}}\equiv\bigotimes_{j=1}^N \ket{\beta_j}$,
where $\ket{\beta_j}\in\{\ket{+},\ket{-}\}$.
Given a fixed $\ket{\boldsymbol{\beta}}$, we can construct the translational invariant state
\begin{equation}\label{translationally invariant states general}
\ket{\boldsymbol{\beta}, p}=\frac{1}{\sqrt{N}}\sum_{j=0}^{N-1}e^{-\imath p j} T^{j}\ket{\boldsymbol{\beta}},
\end{equation}
which is an eigenstate of the operator $T$ with momentum $p$. 
For instance, the states $\ket{s_p}$ in eq.~\eqref{states s p} can be obtained by setting $\ket{\boldsymbol{\beta}}=\ket{+-+-\ldots+-+}=\ket{N}$ and considering that $\ket{j}=(T^\dagger)^j\ket{N}$. We can write then any ground state $\ket{g_p}$ of $H$ with momentum $p$ as
\begin{equation}\label{ground state decomposition}
\ket{g_p}=\sum_{\boldsymbol{\beta}} c_{\boldsymbol{\beta}} \ket{\boldsymbol{\beta},p},
\end{equation}
where the sum is over all the different, and not equivalent by translation, states $\ket{\boldsymbol{\beta}}$, and the normalization implies $\sum_{\boldsymbol{\beta}}|c_{\boldsymbol{\beta}}|^2=1$. 
Here we say that two states, $\ket{\boldsymbol{\beta}_1}$ and $\ket{\boldsymbol{\beta}_2}$, are not equivalent by translation if $\ket{\boldsymbol{\beta}_1}\neq T^k\ket{\boldsymbol{\beta}_2}$ for any integer $k$. 
For instance, the states $\ket{s_p}$ in \eqref{states s p} are given by $c_{\boldsymbol{\beta}}=1$ for $\ket{\boldsymbol{\beta}}=\ket{+-+-\ldots+-+}=\ket{N}$ and $c_{\boldsymbol{\beta}}=0$ for states $\ket{\boldsymbol{\beta}}$ with more than one kink. 

For a small $\lambda$ compared to the energy gap at the classical point, i.e. for $|\lambda| \ll 1$, in the ground state \eqref{ground state decomposition} the contribution of the states $\ket{\boldsymbol{\beta},p}$, and therefore the overlap $c_{\boldsymbol{\beta}}$, is expected to decrease fast with the number of kinks in the state $\ket{\boldsymbol{\beta}}$.

\section{Matrix elements of local operators}
\label{sec:MatrixElements}

To discuss local order we study the possible values of matrix elements of local operators between different contributions in the ground state decomposition \eqref{ground state decomposition}. 
For the sake of simplicity, at first, we will completely neglect the contributions from the states with more than one kink and focus on the one-kink subspace only. 
Afterwards, we shall generalize our results to ground states that are made by combinations of an arbitrary finite number of kinks.

Before starting, let us point out that by local operators we mean all operators having support over a finite range of lattice sites, not scaling with $N$. 
Due to translation invariance, without losing generality we can assume that the operator has support over the first $L$ sites (for some fixed integer $L$).
Moreover, taking into account that Pauli spin operators together with the unit operator provide a basis at a single site, we have that any local operator can always be written as a linear combination of a finite number of monomials in the Pauli operators $\sigma_1^{\alpha_1}\sigma_2^{\alpha_2}\ldots \sigma_L^{\alpha_L}$, where $\alpha_1,\alpha_2,\ldots, \alpha_L\in\{0,x,y,z\}$ and $\sigma_j^0=\mathbf{1\!\!\!1_j}$. 
Thus, we can focus only on monomials in Pauli operators, that either commute or anticommute with a given parity operator. 
The following theorem holds:

\begin{theorem}\label{theorem one kink}
Let $A \equiv \sigma_1^{\alpha_1}\sigma_2^{\alpha_2}\ldots \sigma_L^{\alpha_L}$ be a product of Pauli operators, for some integer $L$. Let us consider two states (not necessarily different) of the form as in eq.~\eqref{states s p}, $\ket{s_{p_1}}$ and $\ket{s_{p_2}}$, and let us consider arbitrary superpositions $\ket{g_j}=(u_j \mathbf{1}+v_j\Pi^z)\ket{s_{p_j}}$, for $j=1,2$, where $|u_j|^2+|v_j|^2=1$. We have:
	\begin{enumerate}[label=\alph*)]
	\item if $A$ is such that $\alpha_j\in\{0,x\}$ for all sites $j\in\{1,2,\ldots,L\}$, with $\alpha_j=x$ for an odd number of sites $j$, then 
	\begin{equation}
	  |\bra{g_1}A\ket{g_2}|\leq \frac{C_1}{N|\cos\frac{p_1-p_2}{2}|}. 	\label{bound 1}
	\end{equation}
	\label{case 1}
	\item if in $A$ there is at least one site $j\in\{1,2\ldots, L\}$ for which $\alpha_j\in\{y,z\}$, then
	\begin{equation}\label{bound 2}
	  |\bra{g_1}A\ket{g_2}|\leq \frac{C_2}{N}.
	\end{equation}	
	\label{case 2}
	\end{enumerate}
	Here $C_1$ and $C_2$ are positive constants independent of $N$.
\end{theorem}

Note that the first term in \eqref{bound 1} is well defined, since, by the quantization of the momenta, with $N$ being odd and finite, we cannot have $p_1-p_2=\pm\pi$.
A formal proof of the theorem is provided in the Appendix~\ref{Appendix Proof}, but its basic argument lays on the fact that for any single-kink-state, apart from the two spins that are aligned (the kink), there is a perfect alternation of eigenstates of $\sigma^x$.

In case \textit{\ref{case 1}}, the operator $A$ commutes with $\Pi^x$ and hence the evaluation of $\bra{g_1}A\ket{g_2}$ reduces to the evaluation of $\bra{s_{p_1}}A\ket{s_{p_2}}$. 
Moreover, by construction, the kink states are also eigenstates of any $\sigma^x_j$, and hence also of $A$ in this case. Thus only matrix elements between the same kink state are different from zero and we have
\begin{equation}
\bra{s_{p_1}}A\ket{s_{p_2}}=\frac{1}{N}\sum_{j=1}^{N} e^{-\imath (p_1-p_2 )j} \bra{j}A\ket{j}.
\label{deltasum}
\end{equation}
For $j>L$, it is easy to see that $\bra{j}A\ket{j}=c(-1)^j$, for some constant $c\in\{-1,1\}$.
The result in eq.~\eqref{bound 1} comes from inserting the above expectation in eq.~\eqref{deltasum} for the whole sum, and bounding the correction due to the first $L$ elements in the sum differing from the rest.

The case \textit{\ref{case 2}} splits in two different sub-cases.
If the number of sites with $\alpha_j\in\{y,z\}$ is even, the operator $A$ still commutes with $\Pi^x$ and thus the evaluation of $\bra{g_1}A\ket{g_2}$ reduces to the evaluation of $\bra{s_{p_1}}A\ket{s_{p_2}}$.
However, now, differently from  \textit{\ref{case 1}} the kink states are no more eigenstates of the operator $A$.
On the contrary, since the operator $A$ flips some spins, it maps a kink state to a different one and hence the matrix elements between the same kink state vanish. 
Moreover, if the kink is outside the support of $A$, the matrix elements $\bra{j}A\ket{l}$ also vanish because of orthogonality. Thus, we have
\begin{eqnarray}
\bra{s_{p_1}}A\ket{s_{p_2}} & = &\frac{1}{N}\sum_{j,l=1}^{N} e^{-\imath (p_1 j -p_2 l )} \bra{j}A\ket{l} \ ,
\end{eqnarray}
where the terms with $L<j,l<N$ vanish and, hence, we are left with, at most, $(L+1)$ terms of order one, suppressed by the overall factor $1/N$.

On the other hand, if the number of sites with $\alpha_j\in\{y,z\}$ is odd, the operator $A$ anticommutes with $\Pi^x$ and hence the evaluation of $\bra{g_1}A\ket{g_2}$ reduces to the evaluation of $\bra{s_{p_1}} A\Pi^z\ket{s_{p_2}}$. 
Analogously to the previous case we recognize that in the sum
\begin{eqnarray}
\bra{s_{p_1}}A\Pi^z\ket{s_{p_2}} & = &\frac{1}{N}\sum_{j,l=1}^{N} e^{-\imath (p_1 j -p_2 l)} \bra{j}A\Pi^z\ket{l},
\end{eqnarray}
there is at most $(L+1)$ non-vanishing terms, which are of order one and are suppressed by an overall factor that scales with the length of the ring.

The theorem can be generalized straightforwardly to the states with more kinks as follows.
\begin{theorem}\label{theorem general}
	Let $A \equiv \sigma_1^{\alpha_1}\sigma_2^{\alpha_2}\ldots \sigma_L^{\alpha_L}$ be a product of Pauli operators, for some integer $L$. Let us consider two states of the type as in eq.~\eqref{translationally invariant states general}, $\ket{\boldsymbol{\beta}_1, p_1}$ and $\ket{\boldsymbol{\beta}_2,p_2}$, with momentum $p_1$ and $p_2$ respectively.
	\begin{enumerate}[label=\alph*)]
		\item Let $A$ be such that $\alpha_j\in\{0,x\}$ for all sites $j\in\{1,2,\ldots,L\}$, with $\alpha_j=x$ for an odd number of sites $j$. If $\ket{\boldsymbol{\beta}_1}$ and $\ket{\boldsymbol{\beta}_2}$ are different, and not equivalent by translation, then
		\begin{equation}
		\bra{\boldsymbol{\beta}_1,p_1}A\ket{\boldsymbol{\beta}_2,p_2}=0.
		\end{equation}
		If $\ket{\boldsymbol{\beta}_1}=\ket{\boldsymbol{\beta}_2}$ then 
		\begin{equation}
		|\bra{\boldsymbol{\beta}_1,p_1}A\ket{\boldsymbol{\beta}_2,p_2}|\leq  \frac{C_1}{N|\cos\frac{p_1-p_2}{2}|}. 	\label{bound 1 general}
		\end{equation}
		\label{case 1 general}
		\item Let $A$ be such that there is at least one site $j\in\{1,2\ldots, L\}$ for which $\alpha_j\in\{y,z\}$. Then
		\begin{equation}\label{bound 2 general}
		|\bra{\boldsymbol{\beta}_1,p_1}A\ket{\boldsymbol{\beta}_2,p_2}|\leq \frac{C_2}{N}.
		\end{equation}	
		\label{case 2 general}
	\end{enumerate}
	Here $C_1$ and $C_2$ are positive constants independent of $N$, that depend only on $L$ and the number of ferromagnetic bonds in the states $\ket{\boldsymbol{\beta}_1}$ and $\ket{\boldsymbol{\beta}_2}$.
\end{theorem}
The proof of Theorem \ref{theorem general} is similar to the one of Theorem \ref{theorem one kink}, but more involved. The details are also given in the Appendices~\ref{Appendix Proof2}.

\section{Local order in the ground state}
\label{sec:LocalOrder}

Based on the previous theorems, we can now move to discuss the local order in the ground state, depending on the ground state momenta. 
The various ground states, labeled by momentum and parity, can be followed from the classical point $\lambda=0$ to a finite $\lambda$, and represented in terms of states with a progressively growing number of domain wall states. 
With generic boundary conditions, at some critical point $\lambda_c\neq 0$ the system will undergo a quantum phase transition, characterized by a change in the ground state properties, as well as by the non-analytic behavior of the ground state energy (density) in the thermodynamic limit~\cite{Sachdev2011}. 
Since this quantity is not sensitive to the choice of boundary conditions or the odd number of lattice sites, the phase transition point cannot be moved by applying FBC. 
However, a system can also cross smaller, non-extensive, discontinuities (boundary phase transitions), such as the one discussed in~\cite{Maric2020CP}, due to a ground-state level crossing, which also mark a change in the ground state order. 
In any case, the order, or lack thereof, being a characteristic property of a phase between critical points, it is sufficient to study it in a small finite interval of $\lambda$ to determine the nature of a given phase.

We make a natural assumption that in the regime $|\lambda|\ll 1$ the behavior of the local order is captured within the subspace spanned by states with a finite, bounded, although arbitrary, number of kinks. 
We note, for example, that the properties of the magnetization in the exactly solvable quantum XY chain can be captured already within the one-kink subspace~\cite{Maric2019,Maric2020CP}. 
We discuss also the contribution of the states with more kinks, since some interactions can involve preferably such states, and show that they do not change the obtained picture about the relation between the ground state momenta and local order.

If the system's ground space is only two-fold degenerate, i.e. if there exist only a particular momentum $p(N)$ (allowing for system size dependence), with the associated ground states $\ket{g_{p(N)}}$ and $\Pi^z\ket{g_{p(N)}}$, the theorems imply that the expectation values of local operators that break a Hamiltonian symmetry are $O(N^{-1})$. 
In particular, they vanish in the thermodynamic limit.

There is a simple intuitive explanation for this result if we look at the expectation value of $\sigma_j^x$. The states $\ket{g_{p(N)}}$ and $\Pi^z\ket{g_{p(N)}}$ have the same eigenvalue of the translation operator $T$. 
If the ground space is only two-fold degenerate, the consequence is that the expectation value of $\sigma_j^x=(T^\dagger)^j\sigma_N^xT^j$ is independent of $j$ in any ground state. The leading interaction in the model being antiferromagnetic, the ferromagnetic order should not survive in the thermodynamic limit, so it vanishes.

The situation becomes more complex if the system admits a larger ground state degeneracy. 
Let us say that the system has $2d$-fold degenerate ground space and denote the ground state momenta by $p_1(N), p_2(N),\ldots p_d(N)$, whose value depends on the system size, and by $p_1^*, p_2^*,\ldots ,p_d^*$ the values at which they tend in the thermodynamic limit. 
Then, unless $p_n^*-p_m^*= \pi$ for some $n$ and $m$, the theorems imply again that there is no local parameters.
On the other hand, if it is the case that $p_n^*-p_m^*= \pi$ for some $n$ and $m$, then we can construct a ground state such as
\begin{equation}
\ket{g(N)}=\frac{1}{\sqrt{2}}(\ket{g_{p_n(N)}}+e^{\imath \theta}\ket{g_{p_m(N)}}),
\end{equation}
for some phase $\theta$, which exhibits a non-zero order parameter. 
To explain this it is sufficient to focus on the one-kink subspace, where $\ket{g_{p(N)}}=\ket{s_{p(N)}}$. 
Applying the procedure as in the proof of the Theorem~\ref{theorem one kink} we find the site-dependent magnetization 
\begin{equation}\label{magnetization}
\!\!\!\!\!\bra{g(N)}\!\sigma_j^x\!\ket{g(N)}\!=\!\frac{\cos[(p_m(\!N\!)\!-\!p_n(\!N\!))j\!\!+\!\!\theta' ]}{N|\cos\frac{p_n(N)-p_m(N)}{2}|}+\!O(N^{-1})\!\!
\end{equation}
where the phase $\theta'$ is related to $\theta$, but its explicit expression is not needed. 
Since $p_n(N)-p_m(N)=\pi+O(N^{-1})$, in the denominator the correction compensates the factor $N$ and produces a nonvanishing value of the magnetization in the thermodynamic limit.
Moreover, in the numerator it forces a slowly varying magnetization profile.  
In fact, while for neighboring sites the magnetization is almost perfectly staggered, over the whole chain the $1/N$ correction adds up so that the amplitude of the order parameter varies and even locally vanishes at some points. 
Thus, the one in eq.~\eqref{magnetization} is not a standard AFM order and the phase $\theta'$ ($\theta$) allows to select the site on which the minimum of the magnetization (or the maximum) is reached.
A nice example of this phenomenology was discussed for the quantum XY chain with two AFM interactions in~\cite{Maric2020CP}. 
There, the model exhibits a four-fold degenerate ground space, with $p_1(N)=-p_2(N)=\frac{\pi}{2}(1+\frac{1}{N}(-1)^{(N+1)/2})$ so from \eqref{magnetization} we get approximately the magnetization $\bra{g}\sigma_j^x\ket{g}= \frac{2}{\pi} (-1)^j\cos \left( \frac{\pi}{N} j +\theta'' \right)$, which was termed {\it incommensurate antiferromagnetic order}.

Finally, we should also remark that it is possible for the ground state degeneracy to depend on the system size and that a finite order parameter can be reached only through a precise sequence of system sizes. 
This is a peculiar phenomenon in the topologically frustrated models that has no counterpart in the unfrustrated ones. 

\section{Applications on a few examples}
\label{sec:Examples}

\subsection{Models with two-body interactions}
\label{sec:2body}

Let us consider models with only two-body interactions, both nearest-neighbor and beyond. The Hamiltonian of such models has to commute with $\Pi^\alpha$ for $\alpha=x,\,y,\,z$, and the term $H_j$ in eq.~\eqref{Hamiltonian AFM} can be written in the form
\begin{equation}\label{Hamiltonian two-body}
H_j= \sum_{l=1}^r\sum_{\alpha=x,y,z} \mu_{l}^\alpha \sigma_{j}^\alpha\sigma_{j+l}^\alpha \,.
\end{equation}
Here $r$ is the maximal distance between directly interacting spins, $\mu_{l}^\alpha$ is the relative strength of each term and considering that the short range Ising term along $x$ is already accounted we assume $\mu_{l}^x=0$.
Also, since the nearest-neighbor term along the x direction has to be the dominant one,  we further set $|\mu_{l}^\alpha|\le1$.

To begin, let us assume that $r=1$ and $|\lambda|\ll 1$.
Under the latter assumption we can diagonalize the Hamiltonian within the one-kink subspace, i.e. we perform the lowest order-perturbation theory, and determine the ground state momentum. 
It is easy to see that in the one-kink subspace the nearest neighbor interactions 
act by translating the kink by two sites, i.e.
\begin{equation}
\sum_{j=1}^{N}\sigma_j^{\alpha}\sigma_{j+1}^\alpha=T^2+(T^\dagger)^2, \quad \alpha=y,z.
\end{equation}

It follows then that the energy of the translationally invariant states $\ket{s_p}$, $\Pi^z\ket{s_p}$, is
\begin{equation}
E_p=-(N-2)+2 \lambda (\mu_1^y+\mu_1^z)\cos(2p).
\end{equation}
If $\lambda(\mu_1^y+\mu_1^z)<0$, the minimum of $E_p$ is reached for $p=0$ and the ground state manifold is two fold degenerate (one state for each sector of a parity operator). 
Hence, applying Theorem~\ref{theorem one kink}, we obtain that  there is no local order in the thermodynamic limit. 
On the other hand, for $\lambda(\mu_1^y+\mu_1^z)>0$ the minimum of the energy is reached at $p=\pi/2$.
However, as we have already noted, due to the quantization rules, $\pi/2$ is not an admissible value of the momentum for any finite length of the chain. 
As a consequence, for each parity the momenta of the ground states are $p_1(N)=-p_2(N)=\frac{\pi}{2}(1+\frac{1}{N}(-1)^{(N+1)/2})$ and the system exhibits the incommensurate AFM order, discussed above.

\begin{figure}
	\includegraphics[width=8.5cm]{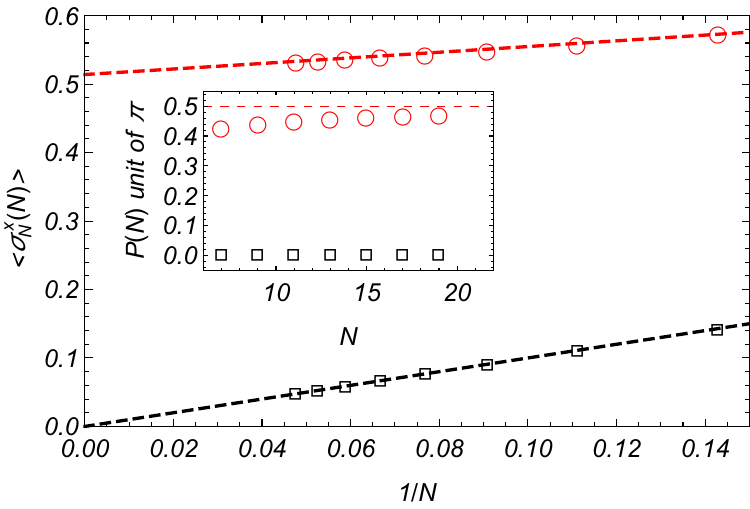}
	\caption{Main Figure: Behavior of the magnetization along the $x$ spin direction at one site
	for the short-range two-body models, as a function of the inverse of the ring length $N$. The data are obtained setting  $\mu^y_1=1/3$ and $\mu^z_1=1/2$, and $\lambda=2/3$ (red empty circles) and $\lambda=-2/3$ (Black empty square). 	
	Inset:  Dependence of the ground-state momenta, in unit of $\pi$, as function of the ring length $N$ for $\mu^y_1=1/3$ and $\mu^z_1=1/2$ and $\lambda=2/3$ (red empty circles) and $\lambda=-2/3$ (Black empty square).}
	\label{first_fig}
\end{figure}

Even if such a picture was obtained in the limit of \mbox{$|\lambda|\ll 1$}, it stands also when finite values of $\lambda$ are considered, as it can be appreciated in Fig.~\ref{first_fig}, where we show results obtained within an exact numerical diagonalization approach.

Going beyond the short-range models, the situation becomes more complex. Roughly speaking, after we have analyzed more than 10.000 realizations of the Hamiltonian in eq.~\eqref{Hamiltonian two-body} with different values of the couplings, we can arrange the various models in two classes. 
The first of these classes is made of models that violate the Quantum Toulouse conditions~\cite{Giampaolo2011,Marzolino13} for an amount that does not scale with the length of the chain, i.e. models in which there is no other source of frustration other than the Topological one induced by FBC.
In such cases, we have that the ground-state manifold is either made by two different elements, and hence no macroscopic order is allowed, or it is four-fold degenerate, and the dependence of the momenta follows the same law of the short-range models, hence allowing for a macroscopic incommensurate order. 
On the contrary, if the Quantum Toulouse conditions are violated for an amount that increases with $N$, other ground-state manifolds are possible, as the one in which the number of independent ground-states depends on the size of the chain, or four-fold degenerate manifolds unable to provide a macroscopic incommensurate order~\cite{Maric2021modify}.

\subsection{Cluster-Ising models}
\label{sec:ClusterIsing}

To provide a specific example of a system where the existence of local order depends on the particular sequence of (odd) system sizes followed towards the thermodynamic limit, we consider the exactly solvable one dimensional $n$-Cluster-Ising models, defined by the Hamiltonian
\begin{equation} \label{Hamiltonian Cluster Ising n}
\!\!H\!=\!\sum_{j=1}^{N}\!\sigma_j^x\sigma_{j+1}^x\!+\!\lambda\!\sum_{j=1}^{N}\!\sigma_j^y(\sigma_{j+1}^z\sigma_{j+2}^z\ldots \sigma_{j+n}^z)\sigma_{j+n+1}^y,
\end{equation}
with $n$ an even number (in order to commute with all the parity operators). 
While the solution of such models, obtained using an exact mapping to free fermions, is known for a few years~\cite{Smacchia2011,Pachos2004,Montes2011,Giampaolo2015,GiampaoloH2015,Zonzo2018}, under FBC a few subtleties have to be taken into account and are presented in the Supplementary Material.

With FBC, we find that the ground state degeneracy of the $n$-Cluster-Ising models depends on the greatest common divisor ($\gcd$) between the system size $N$ and the size $n+2$ of the cluster in the many-body interactions.
In particular, denoting $g\equiv\gcd(N,n+2)$, for $\lambda\in(0,1)$ there are $4g$ ground states, while for $\lambda\in(-1,0)$ the degeneracy is halved (at $\lambda=0$ there is a level crossing, analogous to the one in the XY chain \cite{Maric2020CP}). 
The ground state degeneracy of the topologically frustrated $n$-Cluster-Ising models is thus another example~\cite{Mussardo2020,Mussardo2017,GarciaMartin2020} how the question of divisibility of numbers can appear in quantum mechanics.

A detailed proof of this peculiar behavior of the degeneracy of the ground state manifold can be found in the Appendix~\ref{ClusterIsingApp}.
Here we limit ourselves to a simple and intuitive explanation based on the symmetry of the model.
At the beginning we observe that the Hamiltonian in \eqref{Hamiltonian Cluster Ising n} can be rewritten as $H=\sum_jR_j$ where the operators $R_j$ are 
\begin{equation} \label{RJ operator}
R_j \equiv \sigma_{j-1}^x\sigma_j^x +\lambda \sigma_{j}^y\big( \sigma_{j+1}^z\sigma_{j+2}^z\ldots\sigma_{j+n}^z\big) \sigma_{j+n+1}^y. 
\end{equation}
The $R_j$ operators can be arranged in $g$ terms so that $ H=\sum_{k=1}^{g}H^{(k)}$, 
where each single $H^{(k)}$ is
\begin{equation}
H^{(k)}=\sum_{j=1}^m R_{(j-1)g+k}
\end{equation}
with $m=N/g$.
The different Hamiltonians $H^{(k)}$ mutually commute ($[H^{(k)},H^{(l)}]=0$) and they are invariant under translations by $g$ lattice sites($[H^{(k)},T^g]=0$). 
Due to frustration, the ground state of $H$ cannot minimize the energy of all $H^{(k)}$. 
On the other hand, it can be chosen as a ground state of $g-1$ Hamiltonians $H^{(k)}$ and the first excited state of the remaining one. 
Due to $g$ possible choices of the excited one, the ground state degeneracy of $H$ is at least $g$-fold. 
Since the Hamiltonians $H^{(k)}$ commute with $T^g$ it can be shown that this degeneracy allows for the shift of the momentum by $2\pi/g$ in the ground space: If $p$ is the ground state momentum, so is $p+2\pi/g$.
Furthemore, the mirror symmetry of $H$ (the symmetry under the transformation $\sigma_j^\alpha \to \sigma_{-j}^\alpha$ for $\alpha=x,y,z$ and all $j$) implies that for each ground state with momentum $p$ there is a ground state with momentum $-p$. 

Now, there are two-possible cases. 
The first one is that for any ground state momentum $p$ the momentum $-p$ can be obtained by adding a certain number of increments $2\pi/g$ to $p$.
The second case is that this is not possible. 
In the first case the mirror symmetry does not bring anything new so there are $g$ distinct ground state momenta, while in the second there are $2g$ distinct ground state momenta.
Taking into account also the parity symmetries, it follows that the ground state degeneracy is $2g$ in the first case, and $4g$ in the second. 
It requires the exact solution to see that the first case happens for $\lambda<0$, and the second for $\lambda>0$.

\begin{figure}
	\includegraphics[width=8.5cm]{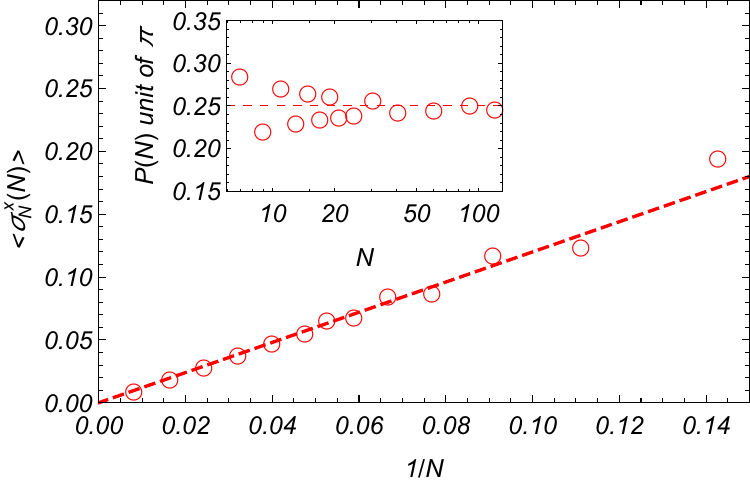}
	\caption{Main Figure: Behavior of the magnetization along the $x$ spin direction at one site
	for the cluster model, as function of the inverse of the ring length $N$. The data are obtained with the analytic approach based on Jordan-Wigner diagonalization for $n=2$ and $\lambda=0.6$. 	
	Inset:  Dependence of the ground-state momenta, in unit of $\pi$, as function of the ring length $N$ for $n=2$ and $\lambda=0.6$.} 
	\label{second_fig}
\end{figure}
Thus, for $n=0,2$ there are $2$ ground states for negative $\lambda$ and $4$ for positive one, for all odd $N$. 
However, these two case are extremely different. 
In fact, assuming $\lambda>0$, while for $n=0$ the 2 distinct ground state momenta tend, in the thermodynamic limit, to $\pm\frac{\pi}{2}$, hence inducing an incommensurate magnetization in the system, for $n=2$ they tend to $\pm\frac{\pi}{4}$ or $\pm\frac{3\pi}{4}$ in the thermodynamic limit. 
As a result, for $n=2$ there is no incommensurate anti-ferromagnetic macroscopic order~\cite{Maric2021modify}, as can be appreciated in Fig~\ref{second_fig}.

\begin{figure}
	\includegraphics[width=8.5cm]{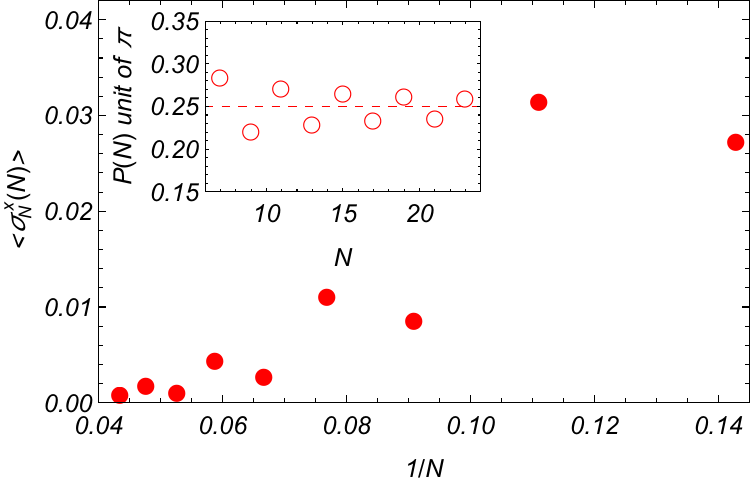}
	\caption{Main Figure: Behavior of the magnetization along the $x$ spin direction at one site
	for a model with next and next-to-near neighbors interactions, as function of the inverse of the ring length $N$. The data are obtained setting  $\lambda=0.8$, $\mu^z_1=\mu_2^y=1/4$ and $\mu^y_1=-\mu_2^x=\mu_2^z=-1/2$. 	
	Inset:  Dependence of the ground-state momenta, in unit of $\pi$, as function of the ring length $N$ for the same model.}
	\label{first_fig_bis}
\end{figure}

Before to go further, it is worth noticing that the same dependence on the momenta from the chain size that characterize the cluster Ising model with $n=2$ can be obtained in systems 
in which $H_j$ obeys to eq.~(\ref{Hamiltonian two-body}) and that violate the quantum Toulouse conditions~\cite{Giampaolo2011}. 
Indeed, in Fig.~\ref{first_fig_bis}, we show the behavior of both the magnetization and the ground-state momenta for the model with  $\mu^z_1=\mu_2^y=1/4$ and $\mu^y_1=-\mu_2^x=\mu_2^z=-1/2$. 
For positive values of $\lambda$ such model shows a four-dimensional ground-state-manifold in which, at each fixed parity, the momenta of the ground-states obey the same rule of the 2-Cluster-Ising-model. 
As a consequence the magnetic order parameters vanish, in the thermodynamic limit, in both systems. 
This fact strongly suggests that, even if the Quantum Toulouse conditions does not apply to models with cluster interactions, these last represent a further source of frustration, that scales with the size of the chain.

\begin{figure}
	\includegraphics[width=8.5cm]{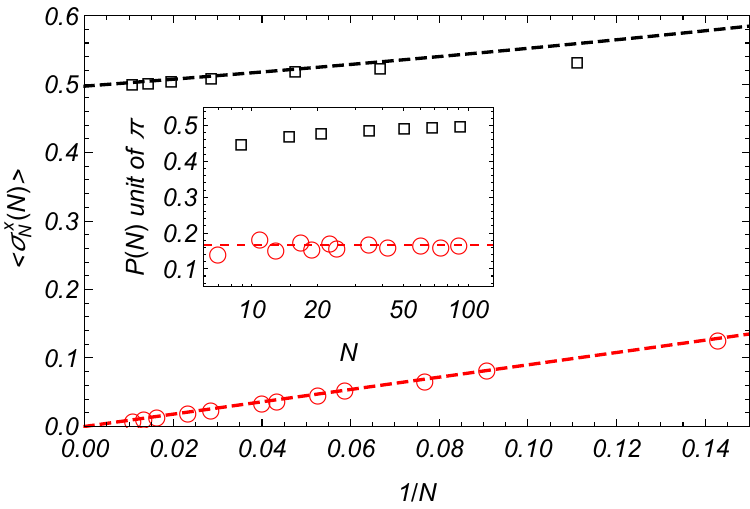}
	\caption{Main Figure: Behavior of the magnetization along the $x$ spin direction at one site
	for the cluster model, as function of the inverse of the ring length $N$. The data, obtained with the analytic approach based on Jordan-Wigner diagonalization for $n=4$ and $\lambda=0.6$, are splitted in two sets: the first for length N=9+6M (empty black squares) and the second for all other odd N (empty red circles). 	
	Inset:  Dependence of the ground-state momenta, in unit of $\pi$, as function of the ring length $N$ for or $n=4$ and $\lambda=0.6$. Also in this case the data are splitted in two sets: the first for length N=9+6M (empty black squares) and the second for all other odd N (empty red circles).} 
	\label{third_fig}
\end{figure}

The situation changes abruptly if we consider $n=4$. 
Let us focus on $\lambda >0$ and take into account separately two chain length sequences, $N=6M+3$ and $N=6M\pm1$ for any positive integer $M$.
Assuming $N=6M\pm1$ the ground space is $4$-fold degenerate and the momenta are $p_1(N)=\frac{\pi}{6} \left( 1 \mp \frac{1}{N} \right)$ and, due to the mirror symmetry, $p_2(N)=-p_1(N)$.
Letting $M\to\infty$ we have $p_1^*-p_2^*=\frac{\pi}{3} \neq \pi$ and thus there is no finite local order parameters in the thermodynamic limit for these chain lengths. 
On the other hand, for $N=6M+3$ the ground space is $12$-fold degenerate, with momenta $p_j(N)=(2j+1)\frac{\pi}{6}-(-1)^j \frac{\pi}{2N}$ for $j=1,2,\ldots,6$. 
In this case we have, for instance, $p_4^*-p_1^*=\pi$ so the system can exhibit a non-zero magnetization. 
From \eqref{magnetization} we find the magnetization $\braket{\sigma_j^x}=\frac{2}{\pi}(-1)^j\cos \left( \frac{\pi}{N} j +\theta \right)$, where the phase factor $\theta$ depends on the ground state choice, see Fig~\ref{third_fig}.

\section{Conclusions}
\label{sec:Conclusions}

In conclusions, we have studied generic Hamiltonians commuting with the three parity operators and examined the expectation values of local operators breaking a Hamiltonian (parity) symmetry.
With a dominant antiferromagnetic Ising interaction and in a setting that induces topological frustration we have shown that there are two possibilities: a) The expectation values of all such local operators decay algebraically, or faster, with the system size and vanish in the thermodynamic limit; b) there is a ground state choice that admits a finite magnetic order, but at the price of breaking the translational invariance. 
Limiting ourselves to models in which the only source of frustration is the topological one induced by boundary conditions, the algebraic decay of the order parameter is associated with the presence of a two-fold degenerate ground state while the presence of an incommensurate order parameter always characterize the four dimensional ground state manifold. 
On the contrary, if other source of frustrations are in the systems, i.e. if the quantum Toulouse conditions are violated for an amount that scales with the system size, we can have other situations, with ground-state vectors with momenta which are incompatible with the existence of a finite incommensurate order parameter. 
In this picture cluster terms, acting simultaneously on an even number of spins, can be seen as a further source of frustration even if quantum Toulouse conditions cannot be applied.
Which of the two possibilities is realized can also depend on the choice of the subsequence of (odd) chain lengths followed towards infinity, as our analysis of the Cluster-Ising models demonstrate. 
We conclude that FBC are special for generic systems: since a perfect AFM order is not compatible with them, either the system disorders or it spontaneously breaks translational symmetry. 
While these findings are probably not robust against a single ferromagnetic defect, we should stress once more that in~\cite{Torre2021} it was shown that the standard AFM order does not reappear in presence of at least one AFM defect, because, following also~\cite{CampostriniRings}, FBC are at the verge of a phase transition and an AFM defect pushes the system into a phase that is either disordered or incommensurate.

These results are intuitive from one side, but very surprising from the point of view that the onset of local order is supposed to be independent from the applied boundary conditions and show once more that frustrated systems (even weakly frustrated ones) belong to a different class of systems altogether.

\section*{Acknowledgments}
We acknowledge support from the European Regional Development Fund -- the Competitiveness and Cohesion Operational Programme (KK.01.1.1.06 -- RBI TWIN SIN) and from the Croatian Science Foundation (HrZZ) Projects No. IP--2016--6--3347 and IP--2019--4--3321.
SMG and FF also acknowledge support from the QuantiXLie Center of Excellence, a project co--financed by the Croatian Government and European Union through the European Regional Development Fund -- the Competitiveness and Cohesion (Grant KK.01.1.1.01.0004).

\begin{appendices}
		
\section{Proof of Theorem 1} 
\label{Appendix Proof}
		
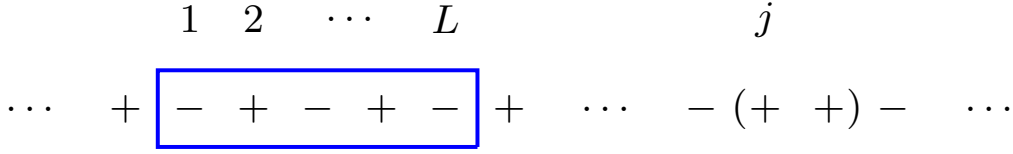
\begin{figure*}[t]
	\begin{tikzpicture}[scale=1.7,every node/.style={scale=1.7}]
		\def \n{7};
		
		\def \f{2};
		
		\draw ({2/\f},0) node[align=center]{$-$};
		\draw ({4/\f},0) node[align=center]{$-$};
		\draw ({6/\f},0) node[align=center]{$-$};
		
		\draw ({1/\f},0) node[align=center]{$+$};
		\draw ({3/\f},0) node[align=center]{$+$};
		\draw ({5/\f},0) node[align=center]{$+$};
		\draw ({7/\f},0) node[align=center]{$+$};
		
		\draw ({(\n+1.5)/\f},0) node[align=center]{$\ldots$};
		
		\draw ({(-0.5)/\f},0) node[align=center]{$\ldots$};
		\draw ({(\n+3)/\f},0) node[align=center]{$-$};
		\draw ({(\n+4)/\f},0) node[align=center]{$+$};
		\draw ({(\n+5)/\f},0) node[align=center]{$+$};
		\draw ({(\n+6)/\f},0) node[align=center]{$-$};
		\draw ({(\n+7.5)/\f},0) node[align=center]{$\ldots$};
		
		\draw ({(\n+3.6)/\f},0) node[align=center]{$($};
		\draw ({(\n+5.4)/\f},0) node[align=center]{$)$};
		
		\def \h{0.7};
		\draw ({2/\f},\h) node[align=center]{$1$};
		\draw ({3/\f},\h) node[align=center]{$2$};
		\draw ({4.5/\f},\h) node[align=center]{$\ldots$};
		\draw ({6/\f},\h) node[align=center]{$L$};
		\draw ({(\n+4)/\f},\h) node[align=center]{$j$};
		
		\def \sh{1.5};
		\def \sv{0.3};
		\def \hors{6.5}
		
		\draw[ blue,line width= 1.5 pt] (\hors/\f,-\sv)--(\sh/\f,-\sv)--(\sh/\f,\sv)--(\hors/\f,\sv)--(\hors/\f,-\sv);
		
	\end{tikzpicture}
	\centering
	\caption{Graphical representation of a kink state $\ket{j}$, where the kink is far on the right. Far from the kink, there is the standard antiferromagnetic order. Let the blue rectangle represent the portion of the lattice where $A$ has the support. Flipping any state in the rectangle will necessarily create a second kink.}
	\label{fig graphical representation}
\end{figure*}

For any $A=\sigma_1^{\alpha_1}\sigma_2^{\alpha_2}\ldots \sigma_L^{\alpha_L}$ we have
\begin{eqnarray}\label{terms}
	\bra{g_1}A\ket{g_2}&=&u_1^*u_2\bra{s_{p_1}}A\ket{s_{p_2}}+u_1^*v_2\bra{s_{p_1}}A\Pi^z\ket{s_{p_2}} \nonumber \\
	&&+v_1^*u_2\bra{s_{p_1}}\Pi^z A\ket{s_{p_2}}
	\nonumber \\
	&&+v_1^* v_2 \bra{s_{p_1}}\Pi^z A \Pi^z \ket{s_{p_2}}. 
\end{eqnarray}
Now, since the kink states $\ket{j}$ are eigenstates of $\Pi^x$, with the eigenvalue $(-1)^{(N-1)/2}$, the states $\ket{s_{p}}$ are also eigenstates of $\Pi^x$, with the same eigenvalue. From this fact and the property that $A$ either commutes or anticommutes with $\Pi^x$ we have that two out of four terms in \eqref{terms} necessarily vanish. If $A$ commutes with $\Pi^x$ ($[A,\Pi^x]=0$) then the second and the third term in \eqref{terms} vanish. Using the Cauchy-Schwarz inequality we get
\begin{equation}\label{bound comm}
	|\bra{g_1}A\ket{g_2}|\leq |\bra{s_{p_1}}A\ket{s_{p_2}}|.
\end{equation}
Similarly, if $A$ anticommutes with $\Pi^x$ ($\{A,\Pi^x\}=0$) then the first and the fourth term vanish and we have
\begin{equation}\label{bound anticomm}
	|\bra{g_1}A\ket{g_2}|\leq |\bra{s_{p_1}}A\Pi^z\ket{s_{p_2}}|.
\end{equation}
Thus we focus our analysis on elements $\bra{s_{p_1}}A\ket{s_{p_2}}$ and $\bra{s_{p_1}}A\Pi^z\ket{s_{p_2}}$. In terms of kink states they read
\begin{align}
	&\bra{s_{p_1}}A\ket{s_{p_2}}=\frac{1}{N}\sum_{j,l=1}^{N} e^{-\imath (p_1j-p_2 l)} \bra{j}A\ket{l}, \label{sum kinks 1} \\
	&\bra{s_{p_1}}A\Pi^z\ket{s_{p_2}}=\frac{1}{N}\sum_{j,l=1}^{N} e^{-\imath (p_1j-p_2 l)} \bra{j}A\Pi^z\ket{l} \label{sum kinks 2}.
\end{align}

\paragraph{Case a):} In case a), $A$ commutes with $\Pi^x$ so \eqref{bound comm} holds. Moreover, $A$ acts only as a phase factor on the kink states, so $\bra{j}A\ket{l}=0$ for $j\neq l$. Since far from the kink we have simply staggered antiferromagnetic order (see Figure \ref{fig graphical representation}) we conclude that for all $j\geq L$ we have
\begin{equation}\label{dependence afm}
	\bra{j}A\ket{j}=c(-1)^j \quad \textrm{for some constant }c\in\{-1,1\}.
\end{equation}
Putting this into \eqref{sum kinks 1} we get
\begin{equation}\label{sum step 1}
	\bra{s_{p_1}}A\ket{s_{p_2}}=\frac{c}{N}\sum_{j=1}^N (-1)^j e^{-\imath(p_1-p_2)j}+\xi_N,
\end{equation} 
where $\xi_N$ is a correction coming from the terms $1\leq j<L$ in \eqref{sum kinks 1} for which \eqref{dependence afm} does not have to hold. It is equal to
\begin{equation}
	\xi_N=\frac{1}{N}\sum_{j=1}^{L-1} e^{-\imath (p_1-p_2)j} \big[ \bra{j}A\ket{j}-c(-1)^j \big]
\end{equation}
and, clearly, satisfies
\begin{equation}
	|\xi_N|\leq \frac{2(L-1)}{N}.
\end{equation}
Performing the sum in \eqref{sum step 1} we are left with
\begin{equation}
	\bra{s_{p_1}}A\ket{s_{p_2}}=-c e^{-\imath (p_1-p_2)/2}\frac{1}{N \cos \frac{p_1-p_2}{2}}+\xi_N.
\end{equation}
Taking the absolute value we get
\begin{equation}
	|\bra{s_{p_1}}A\ket{s_{p_2}}|\leq \frac{1}{N |\cos \frac{p_1-p_2}{2}|}+\frac{2(L-1)}{N}.
\end{equation}
Using \eqref{bound comm} proves this part of the theorem. We can take for the theorem the constant $C_1=1+2(L-1)=2L-1$.

\paragraph{Case b): } The case b) is even simpler. Here $A$ does not act only as a phase on the kink states, but it flips some spins. Flipping any state far from the kink will necessarily create a second kink (see Figure \ref{fig graphical representation}), so all the elements $\bra{j}A\ket{l}$ and $\bra{j}A\Pi^z\ket{l}$ vanish for $L< j < N$ or $L< l<N$. There are thus at most $(L+1)^2$ non-zero elements in the sums \eqref{sum kinks 1} and \eqref{sum kinks 2}. In fact, it's not difficult to give a stronger bound: A kink state $A\ket{l}$, or $A\Pi^z\ket{l}$, can have a non-zero overlap with only one kink-state $\ket{j}$, so there are at most $(L+1)$ non-zero elements in the sum. It follows
\begin{equation}
	|\bra{s_{p_1}}A\ket{s_{p_2}}|\leq \frac{L+1}{N}, \quad |\bra{s_{p_1}}A\Pi^z\ket{s_{p_2}}|\leq \frac{L+1}{N}.
\end{equation}
Now, using \eqref{bound comm} and \eqref{bound anticomm} proves this part of the theorem.

\section{Proof of Theorem 2}
\label{Appendix Proof2}		

\begin{figure*}[t]
	\centering
	\begin{subfigure}{1\textwidth}
		\begin{tikzpicture}[scale=0.5,every node/.style={scale=1.5}]
			
			\begin{scope}[shift={(0,0)}]
				
				\draw (-1.5,0) node[align=center]{$\ket{\boldsymbol{\beta}}=$};
				
				\draw ({0},0) node[align=center]{$|$};
				\draw ({17.5},0) node[align=center]{$\rangle$};

				\draw ({1},0) node[align=center]{$+$};
				\draw ({2},0) node[align=center]{$+$};
				\draw ({3},0) node[align=center]{$+$};
				\draw ({5},0) node[align=center]{$+$};
				\draw ({7},0) node[align=center]{$+$};
				\draw ({9},0) node[align=center]{$+$};
				\draw ({11},0) node[align=center]{$+$};
				\draw ({12},0) node[align=center]{$+$};
				\draw ({14},0) node[align=center]{$+$};
				\draw ({16},0) node[align=center]{$+$};
				
				\draw ({4},0) node[align=center]{$-$};
				\draw ({6},0) node[align=center]{$-$};
				\draw ({8},0) node[align=center]{$-$};
				\draw ({10},0) node[align=center]{$-$};
				\draw ({13},0) node[align=center]{$-$};
				\draw ({15},0) node[align=center]{$-$};
				\draw ({17},0) node[align=center]{$-$};
				
				\def \n{7};
				\draw ({(\n+3.6)},0) node[align=center]{$($};
				\draw ({(\n+5.4)},0) node[align=center]{$)$};
				
				\def \n{1};
				\draw ({(.6)},0) node[align=center]{$($};
				\draw ({(3.4)},0) node[align=center]{$)$};	
				
			\end{scope}

			\begin{scope}[shift={(0,-1.5)}]
				
				\def \sv{0.5};
				
				\def \sh{0.5};
				\def \hors{17.5}
				\draw[line width= 1.5 pt] (\hors,-\sv)--(\sh,-\sv)--(\sh,\sv)--(\hors,\sv)--(\hors,-\sv);
				
				\def \sh{0.5};
				\def \hors{3.5}
				\filldraw[pattern=north east lines] (\hors,-\sv)--(\sh,-\sv)--(\sh,\sv)--(\hors,\sv)--(\hors,-\sv);
				
				\def \sh{10.5};
				\def \hors{12.5}
				\filldraw[pattern=north east lines] (\hors,-\sv)--(\sh,-\sv)--(\sh,\sv)--(\hors,\sv)--(\hors,-\sv);
				
				\draw ({7.},0) node[align=center]{Néel};
				\draw ({15.},0) node[align=center]{Néel};
				
			\end{scope}

		\end{tikzpicture}
		\caption{}
	\end{subfigure}
	
	\vspace{0.5 cm}
	\begin{subfigure}{1\textwidth}
		\begin{tikzpicture}[scale=0.5,every node/.style={scale=1.5}]

			\begin{scope}[shift={(0,0)}]
				
				\draw (-1.5,0) node[align=center]{$\ket{\boldsymbol{\beta}}=$};
				
				\draw ({0},0) node[align=center]{$|$};
				\draw ({17.5},0) node[align=center]{$\rangle$};

				\draw ({1},0) node[align=center]{$+$};
				\draw ({2},0) node[align=center]{$+$};
				\draw ({4},0) node[align=center]{$+$};
				\draw ({6},0) node[align=center]{$+$};
				\draw ({9},0) node[align=center]{$+$};
				\draw ({11},0) node[align=center]{$+$};
				\draw ({13},0) node[align=center]{$+$};
				\draw ({14},0) node[align=center]{$+$};
				\draw ({16},0) node[align=center]{$+$};
				
				\draw ({3},0) node[align=center]{$-$};
				\draw ({5},0) node[align=center]{$-$};
				\draw ({7},0) node[align=center]{$-$};
				\draw ({8},0) node[align=center]{$-$};
				\draw ({10},0) node[align=center]{$-$};
				\draw ({12},0) node[align=center]{$-$};
				\draw ({15},0) node[align=center]{$-$};
				\draw ({17},0) node[align=center]{$-$};
				
				\def \n{0};
				\draw ({(\n+.6)},0) node[align=center]{$($};
				\draw ({(\n+2.4)},0) node[align=center]{$)$};
				
				\def \n{6};
				\draw ({(\n+.6)},0) node[align=center]{$($};
				\draw ({(\n+2.4)},0) node[align=center]{$)$};
				
				\def \n{12};
				\draw ({(\n+.6)},0) node[align=center]{$($};
				\draw ({(\n+2.4)},0) node[align=center]{$)$};
				
			\end{scope}

			\begin{scope}[shift={(0,-1.5)}]
				
				\def \sv{0.5};
				
				\def \sh{0.5};
				\def \hors{17.5}
				\draw[line width= 1.5 pt] (\hors,-\sv)--(\sh,-\sv)--(\sh,\sv)--(\hors,\sv)--(\hors,-\sv);
				
				\def \sh{0.5};
				\def \hors{2.5}
				\filldraw[pattern=north east lines] (\hors,-\sv)--(\sh,-\sv)--(\sh,\sv)--(\hors,\sv)--(\hors,-\sv);
				
				\def \sh{6.5};
				\def \hors{8.5}
				\filldraw[pattern=north east lines] (\hors,-\sv)--(\sh,-\sv)--(\sh,\sv)--(\hors,\sv)--(\hors,-\sv);
				
				\def \sh{12.5};
				\def \hors{14.5}
				\filldraw[pattern=north east lines] (\hors,-\sv)--(\sh,-\sv)--(\sh,\sv)--(\hors,\sv)--(\hors,-\sv);

			\end{scope}

		\end{tikzpicture}
		\caption{}
	\end{subfigure}
	
	\caption{Examples of the first excited states of $H$ at the classical point $\lambda=0$, and their symbolical representation. The white regions represent the Néel order, while the shaded ones include spins that participate in ferromagnetic bonds (kinks).}\label{fig examples excited}
	
\end{figure*}
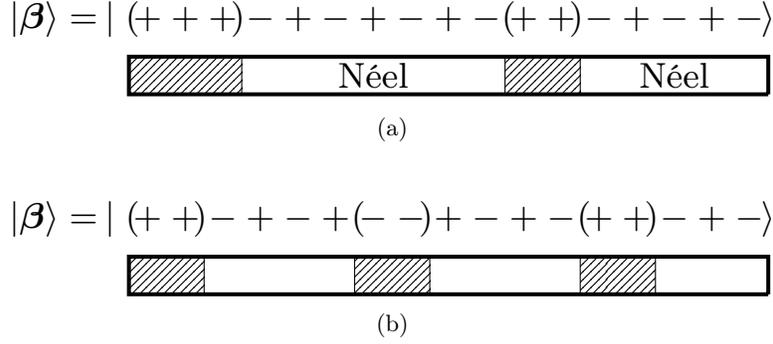

To prove the theorem it is convenient to we write the matrix elements of interest as
\begin{eqnarray}\label{matrix elements more general}
	&& \bra{\boldsymbol{\beta}_1,p_1}A\ket{\boldsymbol{\beta}_2,p_2}=
	\nonumber \\
	&& = \frac{1}{N}\sum_{j,l=0}^{N-1}e^{-\imath (p_1j-p_2l)}\bra{\boldsymbol{\beta}_1}(T^\dagger)^jAT^l \ket{\boldsymbol{\beta}_2}.
\end{eqnarray}
It is also convenient to introduce a symbolical representation of the structure of the states $\ket{\boldsymbol{\beta}}$, in terms of white and shaded regions, as in Figure \ref{fig examples excited}. We define the shaded regions to consist of all spins participating in a ferromagnetic bond (kink) with some of its neighbors, and the white regions to consists of the remaining spins, that participate only in antiferromagnetic bonds. Clearly, the number of white regions in a state $\ket{\boldsymbol{\beta}}$ is equal to the number of shaded regions. Let us denote the number of shaded regions in $\ket{\boldsymbol{\beta}_1}$ and $\ket{\boldsymbol{\beta}_2}$ by $\tilde{N}_1$ and $\tilde{N}_2$ respectively. Let us denote the number of kinks by $N_1$ and $N_2$ respectively. We have then, clearly, $\tilde{N}_1\leq N_1$ and $\tilde{N}_2\leq N_2$. Let us also introduce the concept of the size of a region. We will say that a particular region is of size $R$ if there are $R$ spins inside. For example, in the part a) of Figure \ref{fig examples excited} there are two shaded regions, of size $R_1=3$ and $R_2=2$.

\paragraph{Case a):} In case a) the matrix elements $\bra{\boldsymbol{\beta}_1}(T^\dagger)^jAT^l \ket{\boldsymbol{\beta}_2}$ can be non-zero only if $\boldsymbol{\beta}_1=\boldsymbol{\beta}_2$ and $j=l$, since $A$ acts then only as a phase factor on the eigenstates of $\sigma_j^x$. For $\boldsymbol{\beta}_1\neq\boldsymbol{\beta}_2$ the elements \eqref{matrix elements more general} are thus zero, while for $\boldsymbol{\beta}_1=\boldsymbol{\beta}_2$ we are left with
\begin{eqnarray}\label{matrix elements case a general}
	&&\bra{\boldsymbol{\beta}_1,p_1}A\ket{\boldsymbol{\beta}_1,p_2}
	\nonumber \\
	&& =\frac{1}{N}\sum_{j=0}^{N-1}e^{-\imath (p_1-p_2)j}\bra{\boldsymbol{\beta}_1}(T^\dagger)^jAT^j \ket{\boldsymbol{\beta}_1}.
\end{eqnarray}

Let us focus now on a particular white region in $\ket{\boldsymbol{\beta}_1}$, exhibiting Néel order, and suppose it extends from site $j=r$ to site $j=r+R-1$. This region has a contribution
\begin{equation}
	S\equiv \frac{1}{N}\sum_{j=r}^{r+R-1}e^{-\imath (p_1-p_2)j}\bra{\boldsymbol{\beta}_1}(T^\dagger)^jAT^j \ket{\boldsymbol{\beta}_1}
\end{equation}
in the sum \eqref{matrix elements more general}. If $R>L$ then we have necessarily the staggered dependence
\begin{equation}
	\bra{\boldsymbol{\beta}_1}(T^\dagger)^jAT^j \ket{\boldsymbol{\beta}_1}=c(-1)^j
\end{equation}
for $r\leq j \leq r+R-L$, i.e. before the support of $(T^\dagger)^jAT^j$ starts overlapping with the next shaded region, and with the constant $c\in\{-1,1\}$ given explicitly by $c=(-1)^r\bra{\boldsymbol{\beta}_1}(T^\dagger)^rAT^r\ket{\boldsymbol{\beta}_1}$. We have thus
\begin{equation}\label{step case a general}
	S= \frac{1}{N}\sum_{j=r}^{r+R-1}e^{-\imath (p_1-p_2)j}c(-1)^j+\xi,
\end{equation}
where the correction is given by
\begin{eqnarray}
	\xi & = &=\frac{1}{N}\sum_{j=r+R-L+1}^{r+R-1} e^{-\imath (p_1-p_2)j} \cdot
	\\
	&& \qquad \qquad \qquad \qquad \cdot \big[\bra{\boldsymbol{\beta}_1}(T^\dagger)^jAT^j \ket{\boldsymbol{\beta}_1}-c(-1)^j\big]. \nonumber
\end{eqnarray}
Clearly, the correction satisfies
\begin{equation}
	|\xi|\leq\frac{2(L-1)}{N}.
\end{equation}
Performing the sum in in \eqref{step case a general} we are left with
\begin{equation}
	S=ce^{\imath(p_2-p_1)(r-\frac{1}{2})}\frac{(-1)^{R+1}e^{\imath(p_2-p_1)R}+1}{2N\cos\frac{p_1-p_2}{2}}+\xi.
\end{equation}
Taking the absolute value we get
\begin{equation}\label{white region contribution case a}
	|S|\leq \frac{1}{N|\cos \frac{p_1-p_2}{2}|}+\frac{2(L-1)}{N}.
\end{equation}
In the other case, $R\leq L$, this bound holds trivially. To obtain the bound for the total contribution of the white regions we have to multiply \eqref{white region contribution case a} by the number of white regions in $\ket{\boldsymbol{\beta}_1}$, which is not greater than the number of kinks $N_1$.

We have thus obtained the bound for the contribution of white regions. We can obtain the bound for the contribution of the shaded regions in \eqref{matrix elements case a general} by recognizing that the total number of spins in the shaded regions is not greater than $2N_1$, where $N_1$ is the number of kinks. Altogether, we get
\begin{equation}
	|\bra{\boldsymbol{\beta}_1,p_1}A \ket{\boldsymbol{\beta}_1,p_2}|\leq \frac{N_1}{N|\cos \frac{p_1-p_2}{2}|}+\frac{2(L-1)N_1}{N}+\frac{2N_1}{N},
\end{equation}
so we can take for the theorem the constant
\begin{equation}
	C_1=3N_1+2(L-1)N_1.
\end{equation}

\paragraph{Case b):} We examine the elements $\bra{\boldsymbol{\beta}_1}(T^\dagger)^jAT^l \ket{\boldsymbol{\beta}_2}$ and what are the necessary conditions for them to be nonzero. Let us suppose, without loss of generality, that $\tilde{N}_1\leq\tilde{N_2}$, i.e. the number of shaded regions in $\ket{\boldsymbol{\beta}_2}$ is greater than or equal to the number of shaded regions in $\ket{\boldsymbol{\beta}_1}$.

\begin{figure*}[t]
\begin{subfigure}{0.25\textwidth}
\begin{tikzpicture}[scale=1.,every node/.style={scale=1.5}]

\def \sh{1};
\def \sv{0.5};
\def \hors{3.}
\draw[line width= 1.5 pt,blue] (\hors,-\sv)--(\sh,-\sv)--(\sh,\sv)--(\hors,\sv)--cycle;
\begin{scope}[scale=1.,every node/.style={scale=3.5}]
\draw (3.5,-0.1) node[align=center]{$,$};
\end{scope}

\begin{scope}[shift={(-1,0)}]
\def \sh{0.};
\def \hors{3};
\def \sv{0.3};
\draw[line width= 1.5 pt] (\hors,-\sv)--(\sh,-\sv);
\draw[line width= 1.5 pt] (\hors,\sv)--(\sh,\sv);

\def \sh{1};
\def \sv{0.3};
\def \hors{2.}
\filldraw[pattern=north east lines] (\hors,-\sv)--(\sh,-\sv)--(\sh,\sv)--(\hors,\sv)--cycle;
\draw (0.5,0) node[align=center]{$\ldots$};
\draw (2.5,0) node[align=center]{$\ldots$};
\draw[<->,line width= 1.5 pt] (\sh,0.7)--(\hors,0.7);
\draw (1.5,1.1) node[align=center]{$R_1=3$};
\draw (1.5,-1.37) node[align=center]{ };

\end{scope}	
\end{tikzpicture}
\end{subfigure}
\hspace{ 1 cm}
~
\begin{subfigure}{0.25\textwidth}
\begin{tikzpicture}[scale=1.,every node/.style={scale=1.5}]

\def \sh{1};
\def \sv{0.5};
\def \hors{3.}
\draw[line width= 1.5 pt,blue] (\hors,-\sv)--(\sh,-\sv)--(\sh,\sv)--(\hors,\sv)--cycle;

\begin{scope}[shift={(-0.67,0)}]
\def \sh{0.};
\def \hors{3};
\def \sv{0.3};
\draw[line width= 1.5 pt] (\hors,-\sv)--(\sh,-\sv);
\draw[line width= 1.5 pt] (\hors,\sv)--(\sh,\sv);

\def \sh{1};
\def \sv{0.3};
\def \hors{2.}
\filldraw[pattern=north east lines] (\hors,-\sv)--(\sh,-\sv)--(\sh,\sv)--(\hors,\sv)--cycle;
\draw (0.5,0) node[align=center]{$\ldots$};
\draw (2.5,0) node[align=center]{$\ldots$};	
\end{scope}

\begin{scope}[scale=1.,every node/.style={scale=3.5}]
\draw (3.5,-0.1) node[align=center]{$,$};
\end{scope}

\end{tikzpicture}

\end{subfigure}
\hspace{0.5 cm}
~
\begin{subfigure}{0.25\textwidth}
\begin{tikzpicture}[scale=1.,every node/.style={scale=1.5}]

\def \sh{1};
\def \sv{0.5};
\def \hors{3.}
\draw[line width= 1.5 pt,blue] (\hors,-\sv)--(\sh,-\sv)--(\sh,\sv)--(\hors,\sv)--cycle;

\begin{scope}[shift={(-0.33,0)}]
\def \sh{0.};
\def \hors{3};
\def \sv{0.3};
\draw[line width= 1.5 pt] (\hors,-\sv)--(\sh,-\sv);
\draw[line width= 1.5 pt] (\hors,\sv)--(\sh,\sv);

\def \sh{1};
\def \sv{0.3};
\def \hors{2.}
\filldraw[pattern=north east lines] (\hors,-\sv)--(\sh,-\sv)--(\sh,\sv)--(\hors,\sv)--cycle;
\draw (0.5,0) node[align=center]{$\ldots$};
\draw (2.5,0) node[align=center]{$\ldots$};	
\end{scope}

\begin{scope}[scale=1.,every node/.style={scale=3.5}]
\draw (3.5,-0.1) node[align=center]{$,$};
\end{scope}
\end{tikzpicture}
\end{subfigure}

\vspace{0.3 cm}
\begin{subfigure}{0.25\textwidth}
\begin{tikzpicture}[scale=1.,every node/.style={scale=1.5}]

\def \sh{1};
\def \sv{0.5};
\def \hors{3.}
\draw[line width= 1.5 pt,blue] (\hors,-\sv)--(\sh,-\sv)--(\sh,\sv)--(\hors,\sv)--cycle;

\begin{scope}[shift={(0.,0)}]
\def \sh{0.};
\def \hors{3};
\def \sv{0.3};
\draw[line width= 1.5 pt] (\hors,-\sv)--(\sh,-\sv);
\draw[line width= 1.5 pt] (\hors,\sv)--(\sh,\sv);

\def \sh{1};
\def \sv{0.3};
\def \hors{2.}
\filldraw[pattern=north east lines] (\hors,-\sv)--(\sh,-\sv)--(\sh,\sv)--(\hors,\sv)--cycle;
\draw (0.5,0) node[align=center]{$\ldots$};
\draw (2.5,0) node[align=center]{$\ldots$};	
\end{scope}
\begin{scope}[scale=1.,every node/.style={scale=3.5}]
\draw (3.5,-0.1) node[align=center]{$,$};
\end{scope}
\end{tikzpicture}
\end{subfigure}
~
\begin{subfigure}{0.25\textwidth}
\begin{tikzpicture}[scale=1.,every node/.style={scale=1.5}]

\def \sh{1};
\def \sv{0.5};
\def \hors{3.}
\draw[line width= 1.5 pt,blue] (\hors,-\sv)--(\sh,-\sv)--(\sh,\sv)--(\hors,\sv)--cycle;

\begin{scope}[shift={(0.33,0)}]
\def \sh{0.};
\def \hors{3};
\def \sv{0.3};
\draw[line width= 1.5 pt] (\hors,-\sv)--(\sh,-\sv);
\draw[line width= 1.5 pt] (\hors,\sv)--(\sh,\sv);

\def \sh{1};
\def \sv{0.3};
\def \hors{2.}
\filldraw[pattern=north east lines] (\hors,-\sv)--(\sh,-\sv)--(\sh,\sv)--(\hors,\sv)--cycle;
\draw (0.5,0) node[align=center]{$\ldots$};
\draw (2.5,0) node[align=center]{$\ldots$};	
\end{scope}
\begin{scope}[scale=1.,every node/.style={scale=3.5}]
\draw (4.,-0.1) node[align=center]{$,$};
\end{scope}

\end{tikzpicture}
\end{subfigure}
\hspace{-0.8 cm}
~
\begin{subfigure}{0.1\textwidth}
\begin{tikzpicture}[scale=1.,every node/.style={scale=3}]

\draw (0.,0) node[align=center]{$\ldots$};
\begin{scope}[scale=1.,every node/.style={scale=3.5}]
\draw (1.5,-0.1) node[align=center]{$,$};
\end{scope}
\end{tikzpicture}
\end{subfigure}
\hspace{0.5 cm}
~
\begin{subfigure}{0.25\textwidth}
\begin{tikzpicture}[scale=1.,every node/.style={scale=1.5}]

\def \sh{1};
\def \sv{0.5};
\def \hors{3.}
\draw[line width= 1.5 pt,blue] (\hors,-\sv)--(\sh,-\sv)--(\sh,\sv)--(\hors,\sv)--cycle;

\begin{scope}[shift={(2.,0)}]
\def \sh{0.};
\def \hors{3};
\def \sv{0.3};
\draw[line width= 1.5 pt] (\hors,-\sv)--(\sh,-\sv);
\draw[line width= 1.5 pt] (\hors,\sv)--(\sh,\sv);

\def \sh{1};
\def \sv{0.3};
\def \hors{2.}
\filldraw[pattern=north east lines] (\hors,-\sv)--(\sh,-\sv)--(\sh,\sv)--(\hors,\sv)--cycle;
\draw (0.5,0) node[align=center]{$\ldots$};
\draw (2.5,0) node[align=center]{$\ldots$};	
\end{scope}	
\end{tikzpicture}
\end{subfigure}
\caption{The states $T^l\ket{\boldsymbol{\beta}_2}$ for different values of $l$, where we focus on one shaded region of $\ket{\boldsymbol{\beta}_2}$ and its translations. The blue rectangle represents the sites $1\leq j \leq L$, where the support of $A$ is found. All values of $l$ for which the range $1\leq j \leq L$ either borders or overlaps the shaded region are represented. See the proof of case b) of Theorem \ref{theorem general}.}
\label{fig overlap border}

\end{figure*}
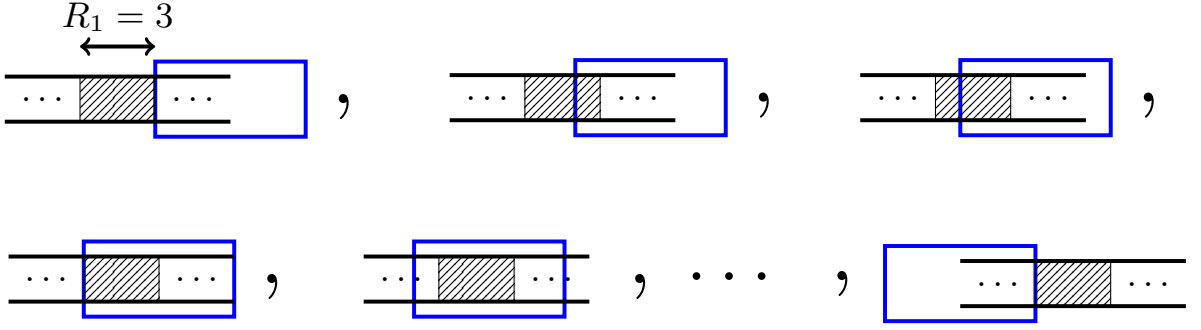

First we notice that all the states $AT^l\ket{\boldsymbol{\beta}_2}$ where $l$ is such that $A$ creates a new shaded region have necessarily zero product with the states $T^j\ket{\boldsymbol{\beta}_1}$, for all $j$, because of a strictly greater number of shaded regions in $AT^l\ket{\boldsymbol{\beta}_2}$ in that case. Thus, first we bound the number of states $T^l\ket{\boldsymbol{\beta}_2}$ in which $A$ does not create a new shaded region. The only states for which this is is a possibility are the states in which the shaded regions overlap or border the range $1\leq j\leq L$, where the support of $A$ is placed, since $A$ creates kinks when acting on a white region. These states are represented in Figure \ref{fig overlap border}.

		For a particular shaded region in $\ket{\boldsymbol{\beta}_2}$, of size $R$, there is at most $R+L+1$ states $T^l\ket{\boldsymbol{\beta}_2}$ which place the shaded region to overlap or border with the the range $1\leq j\leq L$ (see Figure \ref{fig overlap border}). Denoting the sizes of different shaded regions in $\ket{\boldsymbol{\beta}_2}$ by $R_1,R_2,\ldots, R_{\tilde{N}_2}$ we have that there is at most
		\begin{equation}\label{step number bound}
			(R_1+L+1)+(R_2+L+1)+\ldots+(R_{\tilde{N}_2}+L+1)
		\end{equation}
		such states. Recognizing that the total size of the shaded regions is bounded as
		\begin{equation}
			R_1+R_2+\ldots+R_{\tilde{N}_2}\leq 2N_2,
		\end{equation}
		where $N_2$ is the number of kinks in $\ket{\boldsymbol{\beta}_2}$, and that $\tilde{N_2}\leq N_2$, we can bound \eqref{step number bound} by the number $N_2 (L+3)$. Thus, there is at most $N_2 (L+3)$ different values of $l$ for which the product of $AT^l\ket{\boldsymbol{\beta}_2}$ with $T^j\ket{\boldsymbol{\beta}_1}$ is nonzero for some $j$.
		
		The next step is to bound the number of states $T^j\ket{\boldsymbol{\beta}_1}$ which have a nonzero product with a given state $AT^l\ket{\boldsymbol{\beta}_2}$, for fixed $l$. There are two cases to consider. The first one is if there is a shaded region in $T^l\ket{\boldsymbol{\beta}_2}$ that is outside the range $1\leq j \leq L$, or at least a part of size $2$ of the shaded region. In this case the necessary condition for a nonzero value of the elements $\bra{\boldsymbol{\beta}_1}(T^\dagger)^jAT^l \ket{\boldsymbol{\beta}_2}$ is that some shaded region of $T^j\ket{\boldsymbol{\beta}_1}$ coincides exactly with the aforementioned shaded region of $T^l\ket{\boldsymbol{\beta}_2}$. There is at most one such state $T^j\ket{\boldsymbol{\beta}_1}$ for each shaded region of $\ket{\boldsymbol{\beta}_1}$. Thus the number of states $T^j\ket{\boldsymbol{\beta}_1}$ which have a nonzero product with a given state $AT^l\ket{\boldsymbol{\beta}_2}$ is at most $\tilde{N}_1$ in this case. The second case is if in $T^l\ket{\boldsymbol{\beta}_2}$ there is no shaded region, or a part of size $2$, outside the range $1\leq j \leq L$. In this case there is at most $L+1$ states $T^j\ket{\boldsymbol{\beta}_1}$ that give a nonzero product, since the translation of any state by $L+1$ sites will necessarily create a shaded region outside the support of $A$. We can include both cases by taking the sum of the bounds from each one, i.e. the number $L+1+\tilde{N}_1$.
		
		Therefore, there is at most $N_2 (L+3)$ values of $l$ for which give a nonzero product of $AT^l\ket{\boldsymbol{\beta}_2}$ with some of the states $T^j\ket{\boldsymbol{\beta}_1}$ and each of these $N_2 (L+3)$ states has a nonzero product with at most $L+1+\tilde{N}_1$ states $T^j\ket{\boldsymbol{\beta}_1}$. We conclude
		\begin{equation}
			|\bra{\boldsymbol{\beta}_1,p_1}A\ket{\boldsymbol{\beta}_2,p_2}|\leq \frac{(L+1+\tilde{N}_1)(L+3)N_2}{N}.
		\end{equation}
		We prefer to express the bound in terms of the number of kinks $N_1$, which satisfies $N_1\geq\tilde{N}_1$, so we can take for the theorem the constant
		\begin{equation}
			C_2=(L+1+N_1)(L+3)N_2.
		\end{equation}

		\section{Example: Cluster-Ising models}
		\label{ClusterIsingApp}
		To illustrate our results we consider the exactly solvable $n$-Cluster-Ising models, that describe a system made of spin-$\frac{1}{2}$ in which a short-range two-body Ising interaction competes with a cluster term, i.e. an interaction affecting simultaneously $(n+2)$ contiguous spins of the system. On a one-dimensional lattice with periodic boundary conditions, and taking the Ising interaction to favor an AFM alignment, the Hamiltonian of these models reads
		\begin{equation}
			\label{Hamiltonian_n}
			H= \sum\limits_{j=1}^{N} \sigma_j^x\sigma_{j+1}^x +\lambda\sum\limits_{j=1}^{N} \sigma_{j}^y\big( \sigma_{j+1}^z\sigma_{j+2}^z\ldots\sigma_{j+n}^z\big) \sigma_{j+n+1}^y ,
		\end{equation}
		where $\sigma_{j+N}^\alpha = \sigma_j^\alpha$, for $\alpha=x,y,z$. It is known \cite{Smacchia2011,GiampaoloH2015} that the models described by Hamiltonian \eqref{Hamiltonian_n} can be solved through an exact mapping to a system of free fermions, employing the same techniques as in the diagonalization of the quantum XY chain \cite{Franchini17,Lieb1961}, which can be considered the special case $n=0$. The Cluster Ising models with an even number $n$ belong to the symmetry class considered in this work so we focus on them. Moreover, to study topological frustration, as in the main text, we take the system size to be an odd number $N=2M+1$, and we focus on the parameter region $\lambda\in(-1,1)$, where the Ising coupling is larger than (and dominating over) the cluster one.

		\subsection{Diagonalization of the $n$-Cluster-Ising models}\label{app diagonalization}
		
		We are now diagonalizing Hamiltonian \eqref{Hamiltonian_n}, when $n$ is an even number. Let us note that the procedure works for odd $n$ as well, the difference being in the expression for the energy of the $\pi$-mode in \eqref{energies n} later. The Hamiltonian commutes with $\Pi^z$ and we split the diagonalization in two sectors of $\Pi^z$,
		\begin{equation}
			\label{supp_Hamiltonian_2}
			H=\frac{1+\Pi^z}{2}H^+ \frac{1+\Pi^z}{2} + \frac{1-\Pi^z}{2}H^- \frac{1-\Pi^z}{2} \; .
		\end{equation}
		In each sector the Hamiltonian is quadratic in terms of Jordan-Wigner fermions
		\begin{equation}\label{JW}
			c_j=\Big(\bigotimes_{l=1}^{j-1}\sigma_l^z\Big)\frac{\sigma_j^x+\imath\sigma_j^y}{2}, \quad c_j^\dagger=\Big(\bigotimes_{l=1}^{j-1}\sigma_l^z\Big)\frac{\sigma_j^x-\imath\sigma_j^y}{2}.
		\end{equation}
		It reads
		\begin{eqnarray}
			H^\pm &=&-\sum_{j=1}^{N}(c_jc_{j+1}+c_jc_{j+1}^\dagger+\textrm{h.c.} ) \nonumber \\
			&&+ \lambda\sum_{j=1}^{N}(c_{j}c_{j+n+1}-c_{j}c_{j+n+1}^\dagger+\textrm{h.c.}) \ ,
		\end{eqnarray}
		where $c_{j+N}=\mp c_j$ in the sector $\Pi^z=\pm 1$.
		
		The Hamiltonian in each sector is quadratic so it can be brought to a form of free fermions. To achieve this, first $H^\pm$ are written in terms of the Fourier transformed Jordan-Wigner fermions,
		\begin{equation}\label{fourier transformed JW fermions}
			b_q=\frac{1}{\sqrt{N}}\sum\limits_{j=1}^{N} c_j \ e^{-\imath qj} , \quad b_q^\dagger=\frac{1}{\sqrt{N}}\sum\limits_{j=1}^{N} c_j^\dagger \ e^{\imath qj}  ,
		\end{equation}
		for $q\in\Gamma^\pm$, where the two sets of quasi-momenta are given by $\Gamma^-=\{2\pi k/N \}$ and $\Gamma^+=\{2\pi (k+\frac{1}{2})/N \}$ with $k$ running over all integers between $0$ and $N-1$. The Bogoliubov rotation
		\begin{equation}\label{Bogoliubov particles}
			\begin{split}
				&a_q=\cos\theta_q \ b_q + \imath \sin\theta_q \ b_{-q}^\dagger, \quad q\neq0,\pi\\
				&a_{q}=b_q, \quad q=0,\pi 
			\end{split}
		\end{equation}
		with the Bogoliubov angle
		\begin{equation}\label{arctan}
			\theta_{q}=\arctan \frac{|1 + \lambda \ e^{\imath (n+2)q}| -\lambda\cos\big[(n+1)q\big]-\cos q}{-\lambda\sin\big[(n+1)q\big]+\sin q}
		\end{equation}
		then brings $H^\pm$ to a free fermionic form. The Bogoliubov angle also satisfies
		\begin{equation}\label{exp 2 theta}
			e^{\imath 2\theta_{q}}=e^{\imath q}\frac{1+\lambda\ e^{-\imath (n+2)q}}{|1+\lambda\ e^{-\imath (n+2)q}|}.
		\end{equation}
		After these sets of transformations, the original Hamiltonian is mapped into
		\begin{equation}
			\label{supp_Hamiltonian_3}
			H^\pm=\sum\limits_{q\in\Gamma^\pm}^{} \varepsilon_q \left(a_q^\dagger a_q-\frac{1}{2}\right) ,
		\end{equation}
		where the quasi-particle energies are given by
		\begin{align}
			\varepsilon_q  & =2|1 + \lambda \ e^{\imath (n+2)q} 
			\label{energies n}  \\
			 & =2\sqrt{1 +\lambda^2+ 2\lambda\cos \big[(n+2)q\big] }& \,\,\,\,\,\,\,\, \forall \,q\neq 0,\pi,  \nonumber \\
			\varepsilon_0 & =2( 1+\lambda)& \,\,\,\,\,\,\,\,\,\, q=0\in \Gamma^- ,  \nonumber \\ 
			\varepsilon_\pi& = -2(1+\lambda) \nonumber &  \,\,\,\,\,\,\,\,\,\,q=\pi\in \Gamma^+.
		\end{align}
		
		Before proceeding, let us note one technical subtlety in the diagonalization of the model. The Bogoliubov angle $\theta_q$, defined by \eqref{arctan} can become undefined for some modes $q\neq0,\pi$ also point-wise, by fine--tuning of the parameters $n$, $N$, and $\lambda$. This problem can be circumvented by using \eqref{exp 2 theta} to define the Bogoliubov angle and such points can be neglected.
		
		\subsection{Eigenstates construction for the $n$-Cluster-Ising models}\label{app eigenstates}

		The eigenstates of $H$ are formed by applying Bogoliubov fermions creation operators on the vacuum states $\ket{0^\pm}$, which satisfy $a_q\ket{0^\pm}=0$ for $q\in\Gamma^\pm$ and taking care of the parity requirements in \eqref{supp_Hamiltonian_2}. The vacuum states are given by
		\begin{equation}\label{Bogoliubov vacuum}
			\ket{0^\pm}=\prod\limits_{0<q<\pi,\; q\in\Gamma^\pm} \big(\cos\theta_q-\imath\sin\theta_q \ b_q^\dagger b_{-q}^\dagger \big) \ket{0},
		\end{equation}
		where $\ket{0}=\ket{\uparrow\uparrow...\uparrow}$ is the state of all spin up and the vacuum for Jordan-Wigner fermions, satisfying $c_j\ket{0}=0$. The vacuum states $\ket{0^+}$ and $\ket{0^-}$ both have, by construction, parity $\Pi^z=+1$. The parity requirements in \eqref{supp_Hamiltonian_2} imply that the eigenstates of $H$ belonging to the $\Pi^z=-1$ sector are of the form $a_{q_1}^\dagger a_{q_2}^\dagger...a_{q_{m}}^\dagger\ket{0^-}$ with $q_i\in\Gamma^-$ and $m$ odd, while $\Pi^z=+1$  eigenstates are of the same form but with $q_i\in\Gamma^+$, $m$ even and the vacuum $\ket{0^+}$ used. It is important to stress that the total quasi-momentum of these Hamiltonian eigenstates is also the momentum of the states that generates lattice translations, i.e. the action of translation operator $T$ on these states acts as a phase factor
		\begin{equation}
			T=\exp\big( \imath \sum_{q\in\Gamma^\pm}q a_q^\dagger a_q \big).
		\end{equation}
		This follows from Theorem 1 in the Supplementary Material of \cite{Maric2020CP}. The identification of the quasi-momentum from the exact solution with the momentum allows us to draw a connection between the general results of this work and the exactly solvable cluster models. 
		
		Because of the anticommuting parity symmetries of the model, having constructed the states of one sector, say $\Pi^z=-1$, the states of the other sector can be constructed also by applying the parity operator $\Pi^x$ (or $\Pi^y$), as in \cite{Maric2019,Maric2020CP} and as discussed in the main text. Namely, if $\ket{\psi}$ is the eigenstate of $H$ with $\Pi^z=-1$ then $\Pi^x\ket{\psi}$ is also the eigenstate, with the same energy, but with $\Pi^z=+1$.
		
		From our construction of the eigenstates we see, in particular, that the ground states of the model \eqref{Hamiltonian_n} are the states $a_q^\dagger\ket{0^-}$ and $\Pi^x a_q^\dagger\ket{0^-}$ for all momenta $q\in\Gamma^-$ that minimize the energy \eqref{energies n}. Note that in the studied parameter region $\lambda\in(-1,1)$ the energy of every mode $q\in\Gamma^-$ is positive. A consequence of this fact is that the system is gapless, with the energy gap above the ground state closing as $1/N^2$, a phenomenology analogous to that of Refs. \cite{Maric2019,Maric2020CP,Dong2016,Giampaolo2018}. 
		
		Determining the ground states, their momenta and the ground state degeneracy becomes, thus, a matter of finding the modes $q\in\Gamma^-$ with minimal energy. From \eqref{energies n} we see that the modes $q\in\Gamma^-$ with minimal energy are, for $\lambda\in(0,1)$, those that minimize $\cos[(n+2)q]$, and, for $\lambda\in(-1,0)$, those that maximize $\cos[(n+2)q]$. The number of such momenta is given by the theorem which is the subject of the next section. Denoting by $g=\gcd(N,n+2)$ the greatest common divisor of $N$ and $n+2$, from the theorem it follows  that the number of modes minimizing the energy is $2g$ and $g$ for $\lambda\in(0,1)$ and $\lambda\in(-1,0)$ respectively. Taking into account the two-fold degeneracy between different parity sectors, we conclude that the ground state degeneracy is $4g$ and $2g$ for $\lambda\in(0,1)$ and $\lambda\in(-1,0)$ respectively.
		
		Let us determine explicitly some of the ground state momenta. We are going to focus on the example presented in the main text, given by $n=4$ and $\lambda\in(0,1)$. From part b) of the theorem in the next section we see that the ground state momenta $q\in\Gamma^-$ are those that satisfy
		\begin{equation}\label{relation momentum example}
			\cos(6q)=-\cos \left( \pi \frac{g}{N} \right),
		\end{equation}
		where $g=\gcd(N,6)$. We are going to focus on the cases $N\ \mathrm{mod} \ 12=1$ and $N\ \mathrm{mod} \ 12=3$, that illustrate our points in the main text, while the other cases can be treated in an analogous way. In the case $N\ \mathrm{mod} \ 12=1$ we have $g=1$ so the ground space is four-fold degenerate (corresponding to two different momenta). It is easy to see that momenta
		\begin{equation}
			q=\frac{2\pi}{N}\frac{N-1}{12},\; \frac{2\pi}{N}\frac{11N+1}{12}
		\end{equation}
		indeed belong to $\Gamma^-$ and satisfy the relation \eqref{relation momentum example}, and are thus the ground state momenta. In the case $N\ \mathrm{mod} \ 12=3$ we have, on the other hand, $g=3$ so the ground space is $12$-fold degenerate (corresponding to six different momenta). The ground state momenta are in this case
		\begin{eqnarray}
			q& = &\frac{2\pi}{N}\frac{N-3}{12},\; \frac{2\pi}{N}\frac{3N+3}{12},\; \frac{2\pi}{N}\frac{5N-3}{12},\;
			\nonumber\\
			&& \frac{2\pi}{N}\frac{7N+3}{12}, \; \frac{2\pi}{N}\frac{9N-3}{12}, \; \frac{2\pi}{N}\frac{11N+3}{12}.
		\end{eqnarray}

		\subsection{Extremization of the energy spectrum for the $n$-Cluster-Ising models}\label{sec degeneracy}
		
		In this section we prove the following theorem, that enables to find the ground state degeneracy of the $n$-Cluster-Ising model setting $m =n+2$.
		\begin{theorem}\label{tm}
			Let $m$ and $N$ be positive integers, such that $m<N$ and $N$ is odd. Let us denote their greatest common divisor by $g=\gcd(N,m)$. Consider the function $f(j)=\cos \left( \frac{2\pi m}{N}  j \right)$ defined for $j\in\{0,1,\ldots N-1\}$.
			\begin{enumerate}[label=(\alph*)]
				\item\label{part a} The function $f$ has $g$ maxima on the set $\{0,1,\ldots N-1\}$, where the function reaches value $1$.
				\item \label{part b} The function $f$ has $2g$ minima on the set $\{0,1,\ldots N-1\}$, where the function reaches value $-\cos(\pi g/N)$.
			\end{enumerate}
		\end{theorem}
		
		For the proof we will use the concept of a {\it multiset}, i.e. a set in which elements can repeat. Two multisets are equal if they contain the same elements, with the same multiplicities. We define the multiplication of the multiset of numbers by a constant: If $A=\{\alpha: \alpha\in A\}$ is a multiset of (complex) numbers and $c$ a (complex) number we define the multiplication in the obvious way, by multiplying each element of the multiset by $c$,
		\begin{equation}
			cA=\{c\alpha: \alpha\in A\}.
		\end{equation} 
		We also introduce the distance of a number from a set, or a multiset, of numbers. Let $\beta$ be a (complex) number and $A$ a set, or a multiset. Then the distance of $\beta$ from $A$ is 
		\begin{equation}
			d(\beta; A)=\min\{|\alpha-\beta|: \alpha\in A\}.
		\end{equation}
		More generally $\inf$ should be used instead of $\min$, of course, but for our purposes it is going to be the same.
		
		Now we introduce a definition about modular arithmetic and multisets. Suppose we have two multisets of integers, $A$ and $B$, and let $m$ also be an integer. We say that $A=B \Mod{m}$ if
		\begin{equation}
			\{\alpha\ \mathrm{mod} \ m: \alpha\in A\}=\{\beta\ \mathrm{mod} \ m: \beta\in B\},
		\end{equation}
		i.e. if looking at equalities modulo $m$ the elements and multiplicities are the same.
		
		With these notions introduced we can prove the theorem.
		\begin{proof}
			\begin{enumerate}[label=(\alph*)]
				\item
				
				If we expand the domain $j\in\{0,1,\ldots N-1\}$ of function $f$ to real values $j\in\mathbb{R}$ then it is easy to see that the function is maximized for $j\in \frac{N}{m}\mathbb{Z}$, with value $f(j)=1$. Within our restricted domain of integers, the elements $j$ that minimize the function $f(j)$ are simply those that satisfy both $j\in\{0,1,\ldots N-1\}$ and  $j\in \frac{N}{m}\mathbb{Z}$, i.e. those $j\in\{0,1,\ldots N-1\}$ satisfying
				\begin{equation}\label{step distance first}
					d\Big(j; \frac{N}{m}\mathbb{Z}\Big)=0.
				\end{equation}
				Since $0\leq j\leq N-1$ the condition \eqref{step distance first} is equivalent to
				\begin{equation}
					d\Big(j; \frac{N}{m}\{0,1,2\ldots, m-1\}\Big)=0.
				\end{equation}
				Clearly, there are as many minimizing values $j$ as there are integers in the set
				\begin{equation}
					\frac{N}{m}\{0,1,2\ldots, m-1\},
				\end{equation}
				and this number is, further, equal to the number of zeroes in the multiset
				\begin{equation}
					A\equiv \Big\{N l\ \mathrm{mod}\ m: l=\{0,1,2,\ldots m-1\}\Big\}.
				\end{equation}
				
				We proceed by exploring the properties of the multiset $A$. Bringing $N$ out of the multiset we get
				\begin{equation}
					A=N\{0,1,2\ldots,m-1\} \Mod{m}.
				\end{equation}
				Introducing the greatest common divisor of $N$ and $m$, denoted by $g=\gcd(N,m)$, and defining
				\begin{equation}\label{step B g}
					B=g\{0,1,2,\ldots m-1\}
				\end{equation}
				we can write
				\begin{equation}\label{step B g A}
					A=\frac{N}{g} B \Mod{m}.
				\end{equation}
				
				The first step is to show that $B$ consists of repeating blocks, if we look at equalities $(\mathrm{mod} \ m)$. Multiplying with $g$ in \eqref{step B g} we get trivially
				\begin{equation}
					B=\{0,\gcdM,2\gcdM\ldots,(m-1)\gcdM\} \Mod{m}.
				\end{equation}
				But notice
				\begin{equation}
					g(m-1)\stackrel{(\mathrm{mod}\ m)}{=}m-g=\Big(\frac{m}{g}-1\Big)g .
				\end{equation}
				This means that the multiset $B$ consists $\mathrm{mod}\ m$ of repeating blocks
				\begin{equation}\label{step repeating blocks}
					0,g,2g,\ldots, \Big(\frac{m}{g}-1\Big)g.
				\end{equation}
				We know that the number of elements in the multiset is $m$, while we see that the number of elements in the block is $m/g$. We conclude that the total number of blocks that form the multiset $B$ must be $g$.
				
				The next step is to examine each block as a multiset and show that it is unaffected by multiplication by $N/g$, i.e. that
				\begin{eqnarray}
					&& \frac{N}{g}\{0,g,2g,\ldots \Big(\frac{m}{g}-1\Big) g \}=
					\\
					&& = \{0,g,2g,\ldots \Big(\frac{m}{g}-1\Big) g \} \Mod{m}.
					\nonumber
				\end{eqnarray}
				For this purpose, it is sufficient to show that all elements on the left are different. It is simple to see that this is the case by assuming the contrary and reducing to contradiction. We assume, thus, that there are two elements which are equal,
				\begin{equation}\label{step equality assumed}
					\frac{N}{g}(l_1g)=\frac{N}{g}(l_2g) \Mod{m},
				\end{equation}
				for some $l_1,l_2\in\{0,1,\ldots, m/g-1\}$ such that $l_1<l_2$. The assumed equality implies
				\begin{equation}
					N(l_2-l_1)=0 \Mod{m},
				\end{equation}
				so that $N(l_2-l_1)$ is divisible by $m$, and $(l_2-l_1)N/g$ is divisible by $m/g$. But since $l_2-l_1\in\{0,1,\ldots m/g-1\}$ we have that $l_2-l_1$ is not divisible by $m/g$. It follows that $N/g$ must have common divisors with $m/g$, which is in contradiction with the property of $g$ being the greatest common divisor of $N$ and $m$. Thus, we have shown that each block is unaffected by multiplication by $N/g$.
				
				The last step is to conclude from \eqref{step B g A} that the set $A$ consists of $g$ repeating blocks \eqref{step repeating blocks}. In particular, $A$ contains $g$ zeroes, which proves part \ref{part a} of the theorem.

				\item 
				
				If we expand the domain $j\in\{0,1,\ldots N-1\}$ of function $f$ to real values $j\in\mathbb{R}$ then it is easy to see that $f(j)$ is minimized for $j=\frac{N}{2m}(2l+1), \ l \in \mathbb{Z}$, with value $f(j)=-1$. However, since for odd $N$ these values of $j$ are never integers, they do not coincide with our restricted domain $j\in\{0,1,\ldots,N-1\}$, and we have to find how close to these values we can get. The minimum of $f$ is achieved by those values $j\in\{0,1,\ldots,N-1\}$ that minimize the distance
				\begin{eqnarray}\label{step distance b}
					&& d\Big(j; \frac{N}{2m}\{2l+1: l\in\mathbb{Z}\}\Big)=\\
					&& = d\Big(j; \frac{N}{2m}\{2l+1: l\in\{0,1,\ldots,m-1\}\Big),
					\nonumber
				\end{eqnarray}
				where the equality holds since $0\leq j \leq N-1$. To count all $j$ that minimize the distance we take the following approach. Let us denote the minimal distance by $d_{\min}$. We first count how many values of $l\in\{0,1,\ldots, m-1\}$ have
				\begin{equation}\label{step distance b 1}
					d\Big(\frac{N}{2m}(2l+1); \{0,1,\ldots,N-1\}\Big)=d_{\min},
				\end{equation}
				and then for each such minimizing $l$ we count all $j\in\{0,1,\ldots,N-1\}$ with
				\begin{equation}
					\Big|\frac{N}{2m}(2l+1)-j\Big|=d_{\min},
				\end{equation}
				To each such $l$ there can be associated one or two values of $j$, depending on whether $d_{\min}<1/2$ or $d_{\min}=1/2$ respectively. It is easy to see that, since $m<N$, the same value of $j$ cannot be associated to different values of $l$.
				
				We will now use a similar procedure as in part \ref{part a} and explore the multiset
				\begin{eqnarray}
					C & \equiv & \Big\{N (2l+1)\ \mathrm{mod}\ (2m): \nonumber\\
					&& \qquad \qquad \qquad l=\{0,1,2,\ldots m-1\}\Big\},
				\end{eqnarray}
				which determines the distances of interest.
				Bringing $N$ out we have
				\begin{equation}
					C= N\{1,3,\ldots 2m-1\} \Mod{2m} .
				\end{equation}
				Now we introduce the greatest common divisor $g=\gcd(N,m)=\gcd(N,2m)$, where the last equality holds since $N$ is odd, and define the multiset
				\begin{equation}\label{step D g}
					D=g\{1,3,\ldots, 2m-1\}.
				\end{equation}
				Then we can write
				\begin{equation}\label{step D g A}
					C=\frac{N}{g} D \Mod{2m}.
				\end{equation}
				
				The first step is to show that $D$ consists of repeating blocks, if we look at equalities $(\mathrm{mod} \ 2m)$. Multiplying with $g$ in \eqref{step D g} we get trivially
				\begin{equation}
					B=\{\gcdM,3\gcdM\ldots,(2m-1)\gcdM\} \Mod{2m}.
				\end{equation}
				But notice
				\begin{equation}
					g(2m-1)\stackrel{(\mathrm{mod}\ 2m)}{=}2m-g=\Big(2\frac{m}{g}-1\Big)g .
				\end{equation}
				This means that the multiset $D$ consists $(\mathrm{mod}\ 2m)$ of repeating blocks
				\begin{equation}\label{step repeating blocks D}
					g,3g,\ldots, \Big(2\frac{m}{g}-1\Big)g.
				\end{equation}
				We know that the number of elements in the multiset is $m$, while we see that the number of elements in the block is $m/g$. We conclude that the total number of blocks that forms the multiset $D$ must be $g$.
				
				The next step is to examine each block as a multiset and show that it is unaffected by multiplication by $N/g$, i.e. that
				\begin{eqnarray}
					&& \frac{N}{g}\{g,3g,\ldots \Big(2\frac{m}{g}-1\Big) g \}=\\
					&& = \{g,3g,\ldots \Big(2\frac{m}{g}-1\Big) g \} \Mod{2m}.
					\nonumber
				\end{eqnarray}
				Since $N/g$ is odd, for this it is sufficient to show that all elements on the left are different. It is simple to see that this is the case by assuming the contrary and reducing to contradiction. We assume, thus, that there are two elements which are equal,
				\begin{equation}
					\frac{N}{g}(2l_1+1)g=\frac{N}{g}(2l_2+1)g \Mod{2m},
				\end{equation}
				for some $l_1,l_2\in\{0,1,\ldots, m/g-1\}$ such that $l_1<l_2$. The assumed equality implies \eqref{step equality assumed}, and by the same argument as in part \ref{part a} we conclude there is a contradiction. Thus, blocks are unaffected by multiplication by $N/g$. It follows that $C$ consists of $g$ repeating blocks \eqref{step repeating blocks D}
				
				The last step is to conclude from the block structure of $C$ about the number of minima. We look separately at two cases, $g=m$ and $g<m$. In the first case, $g=m$, we have $2m-g=g$ so $C$ consists only of elements $g=m$, implying the distance
				\begin{equation}
					d\Big(\frac{N}{2m}(2l+1); \{0,1,\ldots,N-1\} \Big)=\frac{1}{2}.
				\end{equation}
				for all $l\in\{0,1,\ldots m-1 \}$. For each $l$ there is necessarily $j\in\{0,1\ldots, N-1\}$ such that
				\begin{eqnarray}
					&& \Big|\frac{N}{2m}(2l+1)-j\Big|=\\
					&& =\Big|\frac{N}{2m}(2l+1)-(j+1)\Big|=\frac{1}{2}.
					\nonumber
				\end{eqnarray}
				Counting all corresponding $j$ and $j+1$ it follows that $f$ has $2g$ minima on the set $\{0,1,\ldots N-1\}$.
				
				In the second case, $g<m$, the values $l$ with
				\begin{eqnarray}\label{step two equations}
					&& N(2l+1)\stackrel{(\mathrm{mod} \ 2m)}{=}g \qquad \qquad \textrm{and} \nonumber\\
					&& N(2l+1)\stackrel{(\mathrm{mod} \ 2m)}{=}2m-g
				\end{eqnarray}
				minimize the distance \eqref{step distance b 1}, with
				\begin{equation}
					d_{\min}=\frac{g}{2m}
				\end{equation}
				Since $d_{\min}<1/2$ in this case, for each such $l$ there is only one value $j\in \{0,1,\ldots,N-1\}$ with
				\begin{equation}
					\Big|\frac{N}{2m}(2l+1)-j\Big|=d_{\min}.
				\end{equation}
				Due to block structure of $C$, there is $g$ values of $l$ satisfying the first and $g$ values satisfying the second equation in \eqref{step two equations}. It follows that the number of minima of $f$ is again $2g.$ Both in the case $m=g$ and $m<g$ the value of the minimum is
				\begin{eqnarray}
					&& \cos\bigg[\frac{2\pi m}{N} \bigg(\frac{N(2l+1)}{2m}\pm\frac{g}{2m}\bigg)\bigg]=
					\nonumber\\
					&&=-\cos\Big(\frac{\pi g}{N}\Big).
				\end{eqnarray}
				
			\end{enumerate}
		\end{proof}
		
		In fact, in the proof of part \ref{part a} of the Theorem the property of $N$ being odd was nowhere used, and the same statement holds for the case of even $N$. The part \ref{part b} would be different in the case of even $N$, since then, in general, $N(2l+1)/(2m)$ could achieve integer values and belong to the domain $\{0,1,\ldots , N-1\}$.

		\subsection{Ground state degeneracy from the symmetries}
		
		Here we explain the ground state degeneracy of the $n$-Cluster-Ising chain based on the symmetries, in details. We denote $g\equiv \gcd(N,n+2)$. Let us introduce the short-hand notation
		\begin{equation}
			R_j \equiv \sigma_{j-1}^x\sigma_j^x +\lambda \sigma_{j}^y\big( \sigma_{j+1}^z\sigma_{j+2}^z\ldots\sigma_{j+n}^z\big) \sigma_{j+n+1}^y 
		\end{equation}
		so that $H=\sum_{j=1}^N R_j$. The Hamiltonian can then be decomposed as
		\begin{equation}
			H=\sum_{k=1}^{g}H^{(k)},
		\end{equation}
		where
		\begin{equation}
			H^{(g)}=R_g+R_{2g}+R_{3g}+\ldots +R_{N}
		\end{equation}
		and
		\begin{equation}
			H^{(k)}=\left(T^{\dagger}\right)^k H^{(g)}T^k
		\end{equation}
		for $k=1,2,\ldots,g-1$.
		All the Hamiltonians $H^{(k)}$ commute with $T^g$. Crucially, we find that all these Hamiltonians mutually commute ($[H^{(k)},H^{(l)}]=0$). Since different cluster terms, i.e. $\sigma_{j}^y\big( \sigma_{j+1}^z\sigma_{j+2}^z\ldots\sigma_{j+n}^z\big) \sigma_{j+n+1}^y  $ for different $j$, mutually commute, to show the latter it is sufficient to show that all the terms $\sigma_{j-1}^x\sigma_{j}^x$ appearing in $H^{(k)}$ for $k\neq g$ commute with all the cluster terms $\sigma_{j}^y\big( \sigma_{j+1}^z\sigma_{j+2}^z\ldots\sigma_{j+n}^z\big) \sigma_{j+n+1}^y $ appearing in $H^{(g)}$. This follows simply from the observation that $\sigma_{0}^y\big( \sigma_{1}^z\sigma_{2}^z\ldots\sigma_{n}^z\big) \sigma_{n+1}^y$ commutes with $\sigma_{j-1}^{x}\sigma_{j}^x$ for $j\in\{0,1,2,\ldots,n+2\}-\{0,g,2g,\ldots, n+2\}$, where the minus sign stands for the exclusion.
		
		Thus, different Hamiltonians $H^{(k)}$ mutually commute and they commute with the total Hamiltonian $H$. Moreover all these operators commute with $T^g$. Since all these operators mutually commute they can be diagonalized simultaneously. Suppose then that the state $\ket{\psi}$ is a common eigenstate of $H$, $T^g$ and $H^{(k)}$ for $k=0,1,\ldots g-1$. Due to topological frustration the ground state of $H$ does not coincide with the ground state of $H^{(k)}$ for all $k$, but it is the first excited state for a particular $k$ and the ground state for the other (it is easy to see for $\lambda=0$, while for general $\lambda\in(-1,1)$ this can be seen in the fermionic picture of the exact solution). This implies that the states $T^k\ket{\psi}$ for $k=0,1,\ldots g-1$ are mutually orthogonal and that the ground state manifold is at least $g$-fold degenerate.
		
		We can relate this degeneracy to the momentum shift. Suppose that $\ket{\psi}$ is an eigenstate of $T^g$ with the eigenvalue $e^{\imath g p}$ for some momentum $p\in \frac{2\pi}{N}\mathbb{Z}$. Any eigenvalue of $T^g$ can be written in this form. It follows that the (normalized) state
		\begin{equation}
			\frac{1}{\sqrt{g}}\left[1+e^{-\imath p} T + \left(e^{-\imath p} T \right)^2+ \ldots + \left(e^{-\imath p} T \right)^{g-1} \right]\ket{\psi} \label{eigenstate translation operator}
		\end{equation}
		is an eigenstate of $T$ with the eigenvalue $e^{\imath p}$. However, since the transformation $p\to p +2\pi/g$ does not change the value of $e^{\imath gp}$, the state obtained from \eqref{eigenstate translation operator} by this transformation is also an eigenstate of $T$. Thus, if there is a ground state with momentum $p$, there is also a ground state with momentum $p +2\pi/g$.
		
		The ground state degeneracy of the model, $2g$ or $4g$, now follows from the parity symmetry and the mirror symmetry, as discussed in the main text.

\end{appendices}

\bibliography{bibliography}

\end{document}